%% file: main.tex
\documentclass[a4paper,12pt]{article}
\usepackage[utf8]{inputenc}
\usepackage[T1]{fontenc}
\usepackage[english]{babel}
\usepackage[a4paper,hmargin=2.5cm,vmargin=2.5cm]{geometry}
\usepackage{lmodern}
\usepackage{microtype}
\usepackage{amsmath,amssymb,amsthm,mathtools}
\usepackage{graphicx}
\graphicspath{{./}{figures/}}
\usepackage{subcaption}
\usepackage{float}
\usepackage{booktabs}
\usepackage[authoryear,round]{natbib}
\usepackage{xcolor}
\usepackage[colorlinks=true,linkcolor=blue!45!black,citecolor=blue!45!black,urlcolor=blue!45!black]{hyperref}
\newcommand{\sym}[1]{\rlap{\textsuperscript{#1}}}

\numberwithin{equation}{section}

\title{What Does Deep Hedging Actually Learn?\\ Delta Corrections, Regime Fragility,\\ and Symbolic Distillation}

\author{Kirill Zernikov\thanks{The author is grateful to Professor Vyacheslav Gorovoy for supervision, discussions of the research idea, comments, and help with interpreting the results.}\\
{\small New Economic School, Moscow, Russia}}
\date{\small\textit{May 2026}}

\begin{document}

\maketitle

\begin{abstract}

This paper studies empirical deep hedging for S\&P 500 index options under a local downside-shortfall reward. It moves beyond performance comparison by asking what the learned hedge does, when it fails, and whether it can be made auditable. TD3 agents are compared with a daily-updated Black-Scholes delta hedge on the same option episodes. In walk-forward tests from 2015 to 2023, the agents usually learn a systematic delta haircut relative to Black-Scholes. The correction is explained by spot-implied-volatility co-movement and often improves accumulated reward and terminal downside variance, but it is regime-fragile: 2022 exposes losses in adverse daily states, while 2023 shows that underhedging can raise ordinary variance when option P\&L is spot-dominated and the volatility channel is unusually weak. Symbolic regression distills the neural policies into compact formulas that can be traded out of sample; these formulas preserve much of the reward, downside-variance, and CVaR advantage over Black-Scholes, and sometimes sharpen it, but inherit the same fragility in difficult regimes.
\end{abstract}



\bigskip

\section{Introduction}

Dynamic hedging is transparent in the Black-Scholes-Merton model but more complex in market data. Trading is discrete, volatility moves with the underlying, and a hedger may care more about adverse daily P\&L than about symmetric variance. This paper studies that setting for long S\&P 500 index calls hedged by short positions in the index.

The convention is fixed throughout: the portfolio is long one call and holds $-\delta_t$ units of the underlying over the next hedge interval. A larger reported hedge ratio therefore means a larger short-index hedge. The Black-Scholes benchmark is recomputed at each rebalancing date from the current option state and current market-implied volatility.

The agent is trained under a local downside-shortfall reward. It is penalized for negative daily hedging P\&L, but non-negative daily P\&L is largely unpenalized. The learned hedge should therefore be read as a downside-oriented empirical policy, not as a classical replication or terminal mean-variance hedge.

These choices make the learned hedge interpretable as a correction to a familiar benchmark. The main empirical regularity is underhedging: the agent usually shorts less underlying than Black-Scholes. The intuition is simple: for equity-index calls, index declines often coincide with higher implied volatility, so the call price may fall less than under fixed volatility. A full Black-Scholes short hedge can then be too large under empirical option-price dynamics.

The empirical design is a walk-forward backtest from 2015 to 2023. For each test year $Y$, the TD3 agent is trained on earlier data, selected using year $Y-1$, and evaluated out of sample on year $Y$. The agent and benchmark trade the same option episodes, use the same observed prices and underlying paths, rebalance on the same dates, and are evaluated under the same P\&L accounting and two-stage bootstrap.

The point is not simply that the agent sometimes beats Black-Scholes. This work asks what economic correction the agent learns, when that correction stops helping, and whether the neural policy can be converted into an explicit rule that can be inspected and traded.

Our contributions are threefold. First, we show that empirical deep hedging under a local downside objective learns a systematic delta haircut relative to daily-updated Black-Scholes, and we connect this haircut to spot, implied-volatility, and option-price co-movement. Second, we document regime fragility rather than treating outperformance as a single unconditional fact: the same underhedge can reduce downside dispersion in most years, lose reward in adverse daily states, or raise ordinary variance when the volatility channel is unusually weak. Third, we use symbolic regression as policy distillation, producing compact hedge formulas and trading them out of sample against both Black-Scholes and the neural actor. The formulas preserve, and in some metrics strengthen, the economic signal, but their failures show why interpretability and regime diagnostics are part of the result rather than cosmetic additions.

The rest of the paper is organized as follows. In Section 2 we present the literature review. Section 3 defines the rewards and P\&L convention. Section 4 presents the empirical deep hedging methodology. Section 5 describes the data, walk-forward design, performance metrics, and bootstrap inference. Section 6 reports walk-forward performance. Section 7 analyzes the learned hedge correction. Section 8 examines regime failures. Section 9 studies symbolic distillation. Section 10 reports robustness checks and Section 11 concludes.

\section{Literature Review}

The analysis connects four strands of literature: model-based hedging, reinforcement-learning approaches to hedging, downside-risk objectives, and interpretable policy distillation. The common theme is the tension between transparent hedge rules and empirical objectives that are not captured by a closed-form Greek.

\subsection{Model-Based Hedging and Empirical Hedge Performance}

The benchmark remains the Black-Scholes-Merton delta hedge \citep{black1973pricing}. Its appeal is transparency: under constant volatility, continuous trading, and frictionless markets, the hedge follows from replication. Its limitations are equally familiar: empirical option markets have volatility smiles and skews, stochastic volatility, jumps, discrete rebalancing, and transaction costs. Richer models, including local-volatility \citep{dupire1994pricing,derman1994riding}, stochastic-volatility \citep{heston1993closed}, and jump-based models \citep{naik1993option,tankov2009jump}, address parts of this gap; \citet{bakshi1997empirical} show that stochastic volatility is particularly important for S\&P 500 hedging.

The most relevant strand studies direct corrections to Black-Scholes deltas. \citet{hull2017optimal} show that the implied-volatility Black-Scholes delta need not minimize hedging-error variance because it ignores the empirical relation between spot and implied-volatility moves. \citet{alexander2012regime} study regime-dependent smile-adjusted deltas. The present paper keeps Black-Scholes as the transparent benchmark, but learns the correction nonparametrically from realized hedging rewards.

\subsection{Deep Hedging and Reinforcement Learning}

Deep hedging treats hedging as a sequential optimization problem rather than as the solution of a pricing PDE. \citet{buehler2019deep} formulate hedging under frictions and convex risk measures as a machine-learning problem. \citet{kolm2019dynamic} connect dynamic hedging to reinforcement learning, and \citet{cao2021deep} and \citet{giurca2021delta} study reinforcement-learning hedges under model-based environments and transaction costs. Because the hedge ratio is continuous, deterministic policy-gradient methods \citep{silver2014deterministic}, DDPG \citep{lillicrap2016continuous}, and TD3 \citep{fujimoto2018addressing} are natural tools, though their results require robustness checks \citep{henderson2018key}.

Our empirical implementation closely follows the TD3 setup of \citet{mikkila2023empirical}, who train and test an agent directly on empirical S\&P 500 option data. This model-free empirical setting is the reason for using their design rather than a GBM or Heston simulator. Related work extends deep hedging to rough volatility \citep{horvath2021deep}, compares DRL algorithms and reward designs \citep{pickard2023deep}, studies frictions and market impact \citep{huang2025deep}, and develops risk-sensitive hedging under dynamic expectile or tail-risk criteria \citep{marzban2023deep,peng2024risk}. This paper differs by focusing on what the empirical policy learns under a local downside-shortfall reward and by distilling that policy into explicit formulas.

\subsection{Downside Risk, Shortfall Risk, and Local Rewards}

The reward is asymmetric, so the relevant risk literature is not limited to variance. Lower partial moments measure below-target outcomes \citep{fishburn1977mean,bawa1975optimal}; related asset-pricing and portfolio work emphasizes downside deviation as an alternative to symmetric variance \citep{harlow1989asset,harlow1991asset,sortino1991downside}. In option hedging, shortfall-risk and partial-hedging problems provide the closest theoretical relatives \citep{follmer2000efficient,schulmerich2003local,coleman2003discrete}.

Those classical problems are not the same as the empirical object studied here. They typically derive hedges under a specified market structure or solve discrete optimization problems after imposing a model or scenario design. By contrast, the present policy is learned from historical option episodes without specifying a parametric process for spot, implied volatility, or option prices. The symbolic-distillation step then writes the learned local-shortfall hedge as closed-form formulas. These formulas are effectively closed-form empirical approximations to a nonparametrically learned local-shortfall policy.

The reward used here is a local version of this idea and follows \citet{mikkila2023empirical}. With the main parameter choice, the agent minimizes a discounted sum of negative daily hedging P\&Ls. Terminal downside variance and CVaR are reported later because they are useful diagnostics, but they are not the optimized objective. This distinction explains why accumulated reward, downside variance, and ordinary variance can rank the same hedge differently.

\subsection{Interpretability, Symbolic Regression, and Policy Distillation}

Neural hedges are hard to audit. This matters in hedging, where a policy must be stress-tested and interpreted across regimes, not only traded as a black box. Policy distillation transfers behavior from complex reinforcement-learning agents to simpler representations \citep{rusu2016policy}; programmatic and symbolic approaches seek policies that can be inspected directly \citep{verma2018programmatically,cranmer2020discovering,makke2024interpretable}. Symbolic regression has been used in finance for objects such as implied-volatility surfaces \citep{luo2023symbolic}, but its use as a traded policy-distillation tool for empirical deep hedging remains limited. The contribution here is to evaluate whether compact formulas preserve the economic behavior of a learned hedge out of sample.

\section{Theoretical Framework}

\subsection{Deep Reinforcement Learning with Continuous Action Domain}

Dynamic hedging is naturally sequential. At each trading date the hedger observes the option state and current inventory, chooses the next hedge ratio, and realizes P\&L over the next market move. Reinforcement learning solves this problem by learning a policy from interactions with the hedging environment rather than by specifying a parametric law for option and underlying dynamics \citep{sutton2018reinforcement,franccois2018introduction}.

Formally, the environment is represented as a Markov decision process. At time $t$, the agent observes a state $s_t \in \mathcal{S}$ and selects an action $a_t \in \mathcal{A}$. In the present application, the state contains option characteristics and current hedge inventory, while the action is the next hedge ratio. The transition kernel satisfies
\begin{equation}
    p(s_{t+1}|s_1,a_1,\ldots,s_t,a_t)
    =
    p(s_{t+1}|s_t,a_t).
\end{equation}
This Markov representation is an approximation, since real option markets contain path dependence through volatility dynamics, liquidity, and microstructure. It is nevertheless the working representation for the feedforward architecture used here.

The agent follows a deterministic policy
\begin{equation}
    a_t = \mu(s_t|\phi),
\end{equation}
where $\phi$ denotes the parameters of the actor network. The objective is to maximize the expected discounted return
\begin{equation}
    J(\mu_\phi)
    =
    \mathbb{E}
    \left[
        \sum_{i=0}^{\infty}\gamma^i r_{t+i}
        \,\middle|\, \mu_\phi
    \right],
    \label{eq:rl_objective}
\end{equation}
where $\gamma \in (0,1]$ is the discount factor and $r_{t+i}$ represents the immediate reward. In finite hedging episodes, the sum ends at episode termination or option maturity.

Because the action is continuous, the implementation uses Twin Delayed Deep Deterministic Policy Gradient (TD3), an actor-critic method. A critic network estimates the action value $Q(s,a|\theta)$, and the actor is updated in the direction that raises the critic's value for the chosen hedge. TD3 stabilizes this procedure by training two critics and using the smaller target value,
\begin{equation}
    y_j
    =
    r_j
    +
    \gamma
    \min_{k=1,2}
    Q_k
    \left(
        s_{j+1},
        \mu(s_{j+1}|\phi^-)+\epsilon_j
        \,\middle|\,
        \theta_k^-
    \right).
    \label{eq:td3_target}
\end{equation}
where $\epsilon_j$ is clipped target-action noise. The two critics reduce positive maximization bias, the noise smooths the value surface, and delayed actor updates make policy changes depend on more stable critic estimates.

\subsection{Reward Function and Hedging P\&L}

The reward function determines what kind of hedge the agent is asked to learn. Unlike a quadratic or terminal mean-variance criterion, the objective here is local and asymmetric. It is closest to lower-partial-moment and shortfall-risk ideas \citep{follmer2000efficient,schulmerich2003local,coleman2003discrete}: adverse hedging errors matter, while favorable hedging outcomes are not treated as risk in the same way.

The empirical environment in this paper evaluates a long call option hedged with a short position in the underlying S\&P 500 index level. Let $C_t$ denote the option mid-price, $S_t$ the underlying index level, and $\delta_t$ the hedge ratio chosen at date $t$. The hedge account holds $-\delta_t$ units of the underlying over the interval $(t,t+1]$. Daily hedging P\&L is
\begin{equation}
    \mathrm{PnL}_{t+1}
    =
    C_{t+1}-C_t
    -
    \delta_t(S_{t+1}-S_t)
    -
    cS_t|\delta_t-\delta_{t-1}|,
    \label{eq:pnl_calc}
\end{equation}
where $c$ is the proportional transaction-cost parameter. The Black-Scholes benchmark is evaluated with the same accounting convention, replacing $\delta_t$ with the Black-Scholes delta computed each day from the current observed option state: current index level, strike, time to maturity, risk-free rate, dividend yield, and current market-implied volatility. It does not freeze the implied volatility observed at the start of the episode. Thus the agent and the benchmark differ only in the hedge rule, not in prices, rebalancing dates, transaction costs, or P\&L accounting.

The hedge is defined on the SPX index level reported in the OptionsDX data. Realized hedge P\&L is therefore calculated from changes in the index level in \eqref{eq:pnl_calc}. Dividend yield enters the Black-Scholes benchmark and the forward-moneyness state variable, but it is not added as a separate realized cash-flow term in the index-level hedge account. This convention keeps the P\&L definition aligned with the price series used in the empirical data and avoids mixing the index-level hedge with the cash-flow accounting of a dividend-paying ETF or stock basket.

The exact implemented reward is computed from daily P\&L as:
\begin{equation}
    r_{t+1}
    =
    10
    \left(
        0.03
        +
        \mathrm{PnL}^{(100)}_{t+1}
        -
        \kappa
        \left|
            \mathrm{PnL}^{(100)}_{t+1}
        \right|^\alpha
    \right),
    \label{eq:reward_step}
\end{equation}
where $\mathrm{PnL}^{(100)}_{t+1}$ is daily P\&L multiplied by 100, $\kappa$ is the penalty parameter, and $\alpha$ is the reward exponent. This specification follows the empirical deep-hedging convention of using an absolute-value penalty rather than a squared penalty, which gives a less explosive learning signal than a local quadratic loss \citep{mikkila2023empirical}. In the main empirical specification, however, the parameter choice $\kappa=1$ and $\alpha=1$ gives the reward a sharper interpretation. Let
\[
    X_{t+1} = \mathrm{PnL}^{(100)}_{t+1}.
\]
Then
\begin{equation}
    r_{t+1}
    =
    10
    \left(
        0.03
        +
        X_{t+1}
        -
        |X_{t+1}|
    \right)
    =
    0.3
    -
    20X_{t+1}^- ,
    \qquad
    X_{t+1}^-=\max(-X_{t+1},0).
    \label{eq:reward_shortfall}
\end{equation}
Thus, for fixed episode length, maximizing expected accumulated reward is equivalent up to scale and an additive constant to minimizing
\begin{equation}
    \mathbb{E}
    \left[
        \sum_{t}
        \gamma^t
        \left(
            \mathrm{PnL}^{(100)}_{t+1}
        \right)^-
    \right].
    \label{eq:daily_shortfall_objective}
\end{equation}
The trained agent is therefore a learned local shortfall hedger. It is not a classical terminal mean-variance hedge and does not directly minimize terminal downside semivariance. Terminal downside variance, CVaR, and ordinary variance are reported below as diagnostics, but the optimized criterion is path-wise: a negative daily hedging error is penalized immediately, while a later positive daily P\&L does not symmetrically offset it. This is why accumulated reward, terminal downside risk, and ordinary variance may disagree in the empirical results.

\section{Deep Hedging Methodology}

\subsection{State Vector and Forward Moneyness}

The state vector follows \citet{mikkila2023empirical}, with one modification: the experiments use forward rather than spot moneyness,
\begin{equation}
    \frac{F_t}{K}
    =
    \frac{S_t e^{(r^f_t-q_t)\tau_t}}{K},
    \label{eq:forward_moneyness}
\end{equation}
where $K$ is the strike, $\tau_t$ is time to maturity, $r^f_t$ is the risk-free rate, and $q_t$ is the dividend yield used in the Black-Scholes calculation. The state observed by the actor is
\begin{equation}
    x_t
    =
    \left(
        \frac{F_t}{K},
        \tau_t,
        h_t,
        \sigma^{\mathrm{IV}}_t
    \right),
    \label{eq:state_vector}
\end{equation}
where $h_t=-\delta_{t-1}$ is the current underlying inventory and $\sigma^{\mathrm{IV}}_t$ is Black-Scholes implied volatility.

Forward moneyness incorporates carry without adding interest rates as a fifth network input. The state representation therefore remains close to the original \citet{mikkila2023empirical} four-input architecture while accounting for the higher-rate environment in the later sample.

Importantly, analytical Greeks are not included in the state vector. If the policy resembles a delta hedge or a correction to it, that structure is learned rather than supplied as an input.

\subsection{Network Architecture and Training Procedure}

The actor and critic architectures are kept close to \citet{mikkila2023empirical}. Inputs are standardized using training-sample moments. The actor has two hidden layers with 256 units each and outputs a continuous hedge ratio in the admissible range. The critics take the state and action as inputs. Hidden layers use leaky ReLU activations, and the networks are optimized with Adam.

Training uses experience replay: each transition $(s_t,a_t,r_{t+1},s_{t+1})$ is stored in a buffer, and critic updates use random minibatches. Exploration is generated by Gaussian noise added to the actor output during training.

Each walk-forward model is trained for 20,000 sampled training episodes. A training episode draws an eligible starting point from the current training window and follows the corresponding 21-day option path. The fixed budget keeps computational effort comparable across years.

\section{Data and Experimental Design}

\subsection{Options Data}
\label{sec:options_data}

The empirical analysis uses S\&P 500 index option chains from OptionsDX.\footnote{\href{https://www.optionsdx.com/product/spx-option-chain/}{OptionsDX SPX option-chain data}.} Each quote contains the underlying index level, strike, expiration date, option type, bid and ask quotes, bid and ask sizes, implied volatility, and Greeks supplied by the data provider. Option prices are measured by mid-quotes. The risk-free rate is the one-year point of the U.S. Treasury yield curve, obtained from the U.S. Department of the Treasury.\footnote{\href{https://home.treasury.gov/policy-issues/financing-the-government/interest-rate-statistics?data=yield}{U.S. Treasury yield curve data}.} The dividend-yield input is used in the Black-Scholes benchmark and in forward moneyness. Code and exact implementation details are publicly available online.\footnote{\href{https://github.com/Kirill-ZG/Interpretable-Empirical-Deep-Hedging}{GitHub repository for code and implementation details}.}

The cleaning procedure removes observations with missing essential fields, invalid prices or strikes, negative bid-ask spreads, and records outside the admissible hedging region. For validation and testing, starting options are selected around strike-to-spot targets
$K/S \in \{0.85,0.90,\ldots,1.15\}$ and maturity targets from 30 to 90 calendar days. Over the 21-day hedging episode, remaining maturity then declines mechanically.

Usable option data are uneven across calendar time. After cleaning and episode construction, valid test episodes rise from 75 in 2015 to 759 in 2022, then fall to 632 in 2023. Because training episodes are sampled uniformly from eligible starting points, later years with richer option panels receive more weight as the walk-forward process expands. Year-by-year results are therefore genuine out-of-sample regime comparisons, but not balanced-panel comparisons.

The P\&L calculation uses mid-prices; transaction costs are deliberately set to $0$ to isolate model effects from frictions. The same option mid-prices and index levels are used for the agent and the Black-Scholes benchmark.

\subsection{Walk-Forward Design}

The empirical experiment is organized as a walk-forward backtest. For a test year $Y$, the agent is trained on years up to $Y-2$, validated on year $Y-1$, and tested out of sample on year $Y$. The validation year is used for model selection. The test year is used only for final evaluation. This design respects calendar time and avoids look-ahead bias.

Because the option surface and the spot-volatility relation change over time, each test-year result corresponds to the model that would have been available before that year began.

\subsection{Episode Construction}

An episode is a 21-trading-day hedging path for one option. The state is updated daily, the hedge is rebalanced daily, and terminal episode P\&L is the sum of daily P\&Ls. A valid validation or testing episode is therefore identified by both an option contract and a 21-day calendar window.

Training episodes may overlap in calendar time. Validation and testing are filtered so that 21-day calendar windows do not partially overlap. Within a selected window, several options may still be present; they share the same underlying path but differ in strike, maturity, implied volatility, and option-price response. The bootstrap procedure below respects both the time-series and cross-sectional dimensions.

\subsection{Performance Metrics}

The primary evaluation object is the distribution of terminal episode P\&Ls. For each test year and strategy, the analysis computes mean terminal P\&L, accumulated reward, 5\% conditional value at risk, downside variance, ordinary variance, and the hedging-defect metric used later.

Variance-based comparisons are reported as log ratios relative to the Black-Scholes benchmark:
\begin{equation}
    \log
    \left(
        \frac{
            \mathrm{Var}(\mathrm{PnL}^{A})
        }{
            \mathrm{Var}(\mathrm{PnL}^{BS})
        }
    \right),
    \label{eq:var_ratio}
\end{equation}
with an analogous construction for downside variance. A positive value means that the agent has higher terminal P\&L variance than the benchmark; a negative value means that the agent has lower variance.

Accumulated reward is reported separately because it corresponds to the training objective. Terminal variance instead measures symmetric dispersion of final hedging outcomes.

\subsection{Bootstrap Inference}

Inference is conducted separately by test year. The bootstrap follows the structure of the validation and testing sample. First, 21-day calendar windows are resampled. Second, option episodes within each selected window are resampled. The first stage accounts for the common spot and volatility path within a calendar window. The second accounts for option-level variation within that path. Thus, this two-stage procedure accounts for variation across market windows and for variation across options exposed to the same market path.

\section{Empirical Performance: Walk-Forward Results}

The empirical results are reported year by year. Each estimate compares the agent with the Black-Scholes benchmark on the same option episodes. Positive values are favorable for accumulated reward, CVaR, and mean terminal P\&L. Negative values are favorable for variance ratios. Statistical significance is assessed using the two-stage bootstrap.

The main message is that the agent's advantage is asymmetric. It is clearest in downside-oriented metrics, while ordinary variance gives a more cautious assessment.

\begin{figure}[H]
    \centering
    \includegraphics[width=\textwidth]{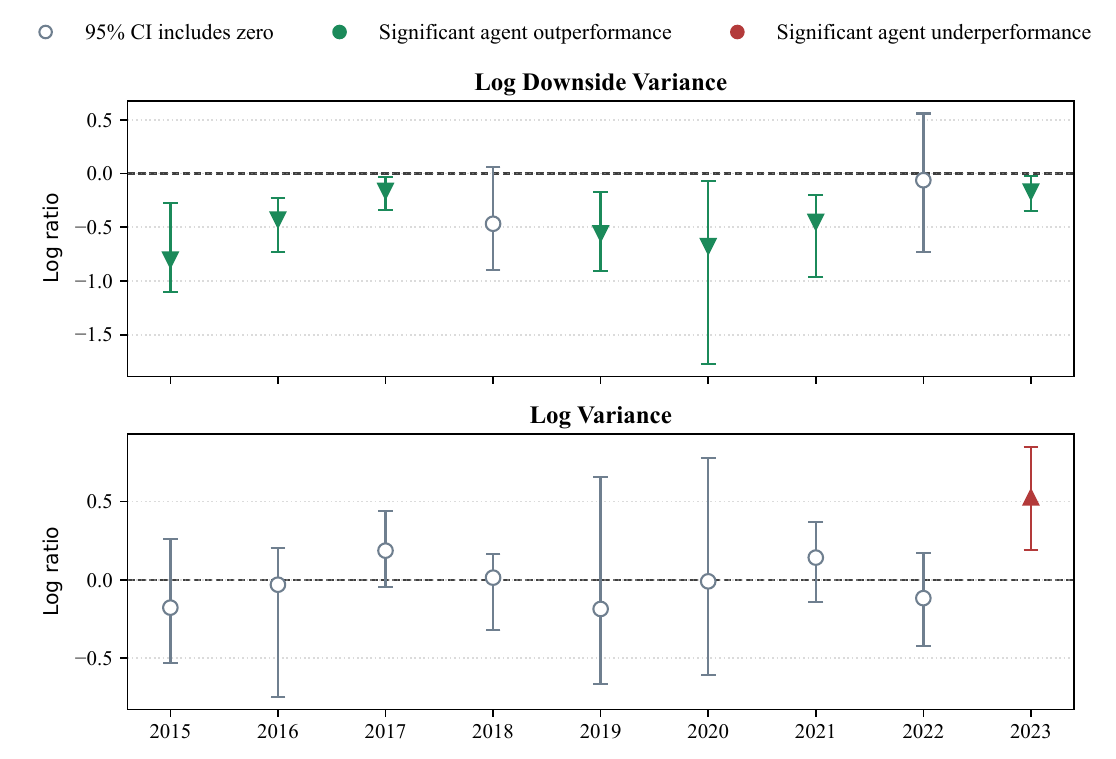}
    \caption{Walk-forward comparison of terminal P\&L downside variance and ordinary variance. Points show agent performance relative to Black-Scholes; bars are two-stage bootstrap 95\% confidence intervals. Negative values are favorable in both panels.}
    \label{fig:wf_variance_downside_pair}
\end{figure}

Figure \ref{fig:wf_variance_downside_pair} shows the central pattern. The top panel is closest to the downside motivation of the reward: terminal downside variance falls significantly in 2015, 2016, 2017, 2019, 2020, 2021, and 2023. The lower panel is more mixed. Ordinary variance is insignificant in most years and significantly higher in 2023. The same policy can therefore compress adverse terminal outcomes while leaving favorable or symmetric dispersion larger. This is not a contradiction; it is the expected tension when a hedge is trained to care more about losses than about all deviations from the mean.

\begin{figure}[H]
    \centering
    \includegraphics[width=\textwidth]{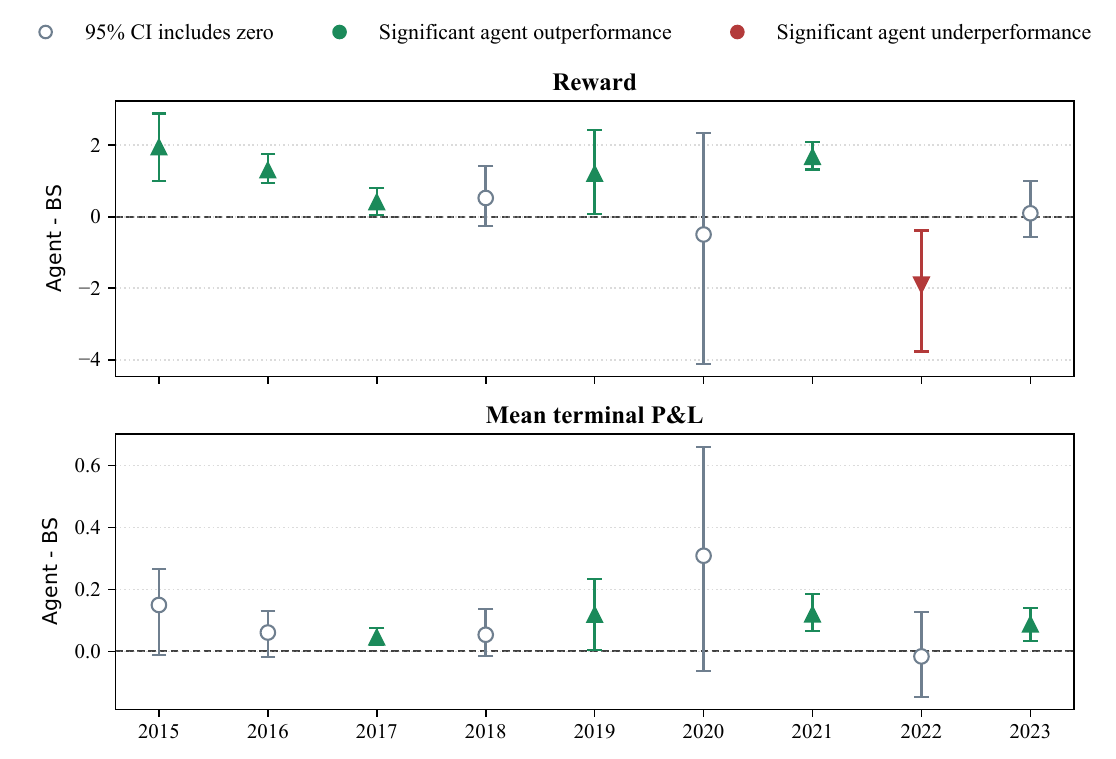}
    \caption{Walk-forward comparison of accumulated reward and mean terminal P\&L. Points show agent performance relative to Black-Scholes; bars are two-stage bootstrap 95\% confidence intervals. Positive values are favorable in both panels.}
    \label{fig:wf_reward_pnl_pair}
\end{figure}

Figure \ref{fig:wf_reward_pnl_pair} reports reward and mean terminal P\&L. Reward is the metric most closely aligned with training, because it aggregates daily shortfall penalties. It improves significantly in five years and fails significantly only in 2022. Mean terminal P\&L is positive in eight of nine years, which is consistent with the interpretation that the learned underhedge often leaves economically favorable residual exposure. The contrast also gives the first warning sign: terminal averages can look acceptable even when the local downside reward deteriorates, because the reward depends on the timing and sign of daily hedging errors rather than only on the final episode P\&L.

Table \ref{tab:wf_metric_summary} puts the four evaluation criteria side by side with the addition of CVaR. The table is useful because no single metric fully describes the learned hedge. Reward measures the optimized local objective; mean terminal P\&L summarizes the average terminal difference; downside variance and CVaR summarize adverse terminal outcomes; ordinary variance penalizes both bad and good outliers. The metric-specific failures are therefore informative rather than incidental. In 2022 the agent loses on accumulated reward without a significant variance loss, pointing to adverse daily shortfall states. In 2023 it improves downside variance but loses on ordinary variance, pointing to favorable residual exposure that raises symmetric dispersion. Section \ref{sec:regime_fragility} explains these cases and the milder 2017 analogue of 2023 variance failure.

\begin{table}[H]
    \centering
    \caption{Compact walk-forward performance summary. Entries are agent-minus-Black-Scholes differences for reward, CVaR, and mean terminal P\&L, and log agent-to-Black-Scholes ratios for downside variance and ordinary variance. $^{*}$, $^{**}$, and $^{***}$ denote 10\%, 5\%, and 1\% two-sided two-stage bootstrap significance, respectively. Positive values are favorable for reward, CVaR, and mean P\&L; negative values are favorable for both variance ratios.}
    \label{tab:wf_metric_summary}
    \input{tables/wf_metric_summary.tex}
\end{table}

\section{The Learned Hedge: A Downside-Oriented Delta Correction}

This section asks what the policy actually does. The main regularity is simple: the learned hedge is usually below the Black-Scholes delta. Because the portfolio is long one call and short the underlying, a negative Agent--BS value in Table \ref{tab:learned_delta_summary} means that the agent shorts less underlying than the Black-Scholes hedge.

\begin{table}[H]
    \centering
    \caption{Average hedge ratios in the walk-forward test years. Agent Delta is the positive hedge ratio implied by the traded agent position; BS Delta is the corresponding Black-Scholes hedge ratio. Underhedged Share is the fraction of hedge intervals in which Agent Delta is below BS Delta.}
    \label{tab:learned_delta_summary}
    \input{tables/learned_delta_summary.tex}
\end{table}

The pattern is persistent. The mean Agent--BS delta gap is negative in every test year, from $-0.009$ in 2017 to $-0.052$ in 2020. The underhedged share exceeds one half in every year and exceeds 75\% in seven of nine years. The network has therefore learned a systematic correction rather than a noisy perturbation around Black-Scholes.

\begin{figure}[H]
    \centering
    \includegraphics[width=\textwidth]{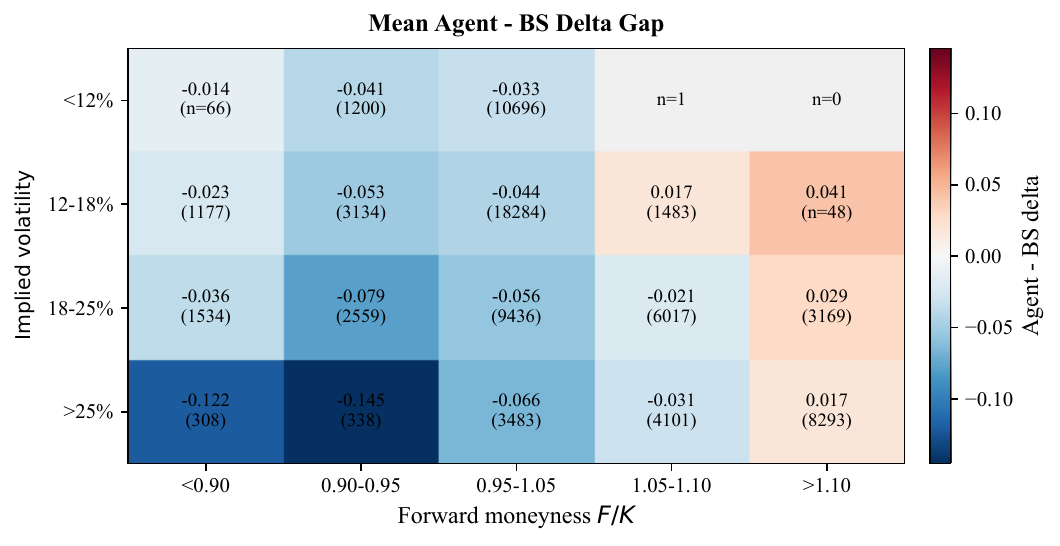}
    \caption{Average Agent--BS delta gap by forward moneyness and implied volatility, pooling all out-of-sample hedge intervals from the main walk-forward agents over 2015--2023. Negative values indicate underhedging relative to Black-Scholes. Parentheses report interval counts; grey cells with zero or one interval are excluded from the heatmap color scale.}
    \label{fig:learned_delta_gap_surface}
\end{figure}

Figure \ref{fig:learned_delta_gap_surface} reports the correction across forward moneyness and implied volatility bins. The main empirical region is negative: the agent generally shorts less underlying than Black-Scholes for out-of-the-money, near-the-money, and moderately in-the-money calls. The correction is strongest in higher-IV and out-of-the-money regions. Deep in-the-money bins can be positive, so the claim is not universal underhedging; it is a broad state-dependent downward correction.

The interpretation is an asymmetric one-instrument hedge, not replication. It leaves positive residual exposure to favorable spot-up moves and reduces losses on the short hedge when the index and call price rise together. The cost is weaker protection when the index falls and the call also loses value. The value of the trade-off depends on the joint behavior of spot, implied volatility, and option prices.

Table \ref{tab:negative_pnl_state_summary} makes the trade-off concrete by grouping negative daily P\&L observations by the directions of spot, option-price, and implied-volatility moves. The table is diagnostic rather than causal: implied volatility is inferred from option prices, so the labels describe realized co-movement.

\begin{table}[H]
    \centering
    \caption{Dominant sources of negative daily P\&L by realized state of the day. S denotes spot and Opt denotes the option price. Loss shares are computed within each portfolio from negative daily P\&L observations only.}
    \label{tab:negative_pnl_state_summary}
    \small
    \input{tables/negative_pnl_ranked_summary.tex}
\end{table}

The two largest rows show the trade-off. The largest share of the agent's negative daily P\&L comes from spot-down, option-down, IV-down states, where underhedging is costly because the volatility channel does not offset the call loss. The largest share of Black-Scholes negative daily P\&L comes from spot-up, option-up, IV-down states, where the short hedge gives back part of the call gain. The learned hedge therefore reallocates exposure rather than eliminating risk.

\section{Regime Fragility of the Learned Correction}
\label{sec:regime_fragility}

The learned correction is useful only when the option surface supports it. This section studies three diagnostic years: 2022, where adverse daily states punish the underhedge; 2023, where Black-Scholes becomes an unusually tight variance hedge; and 2017, a milder low-volatility analogue. All diagnostics are computed on traded walk-forward test episodes and should be read as reduced-form evidence, not structural causal estimates.

\subsection{A Step-Wise Reward Failure in 2022}

The 2022 failure is first a reward failure. The agent loses $1.905$ accumulated reward units relative to Black-Scholes, while neither ordinary nor downside variance shows a significant loss. Since the reward is booked daily, the relevant question is where negative daily P\&L occurs. Under $\kappa=1,\alpha=1$, \eqref{eq:reward_step} is flat for non-negative P\&L and linear for negative P\&L:
\[
    R_{t+1} = 0.3 \quad \text{if } \mathrm{PnL}^{(100)}_{t+1} \geq 0,
    \qquad
    R_{t+1} = 0.3 + 20\,\mathrm{PnL}^{(100)}_{t+1}
    \quad \text{if } \mathrm{PnL}^{(100)}_{t+1} < 0.
\]

\begin{figure}[H]
    \centering
    \includegraphics[width=\textwidth]{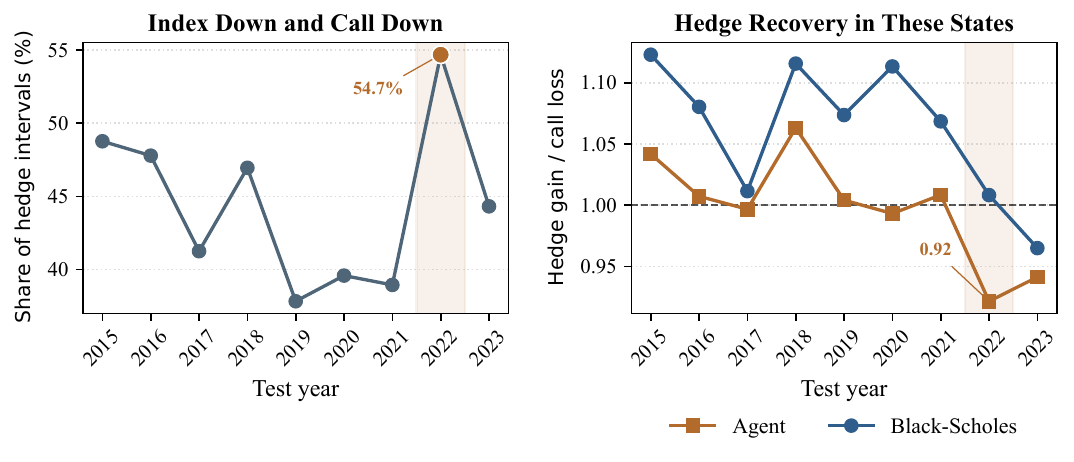}
    \caption{Bad down-state frequency and hedge recovery in the 2022 reward failure. The left panel reports the share of hedge intervals in which the index falls and the call option also loses value. The right panel reports, in those states, mean underlying-hedge gain divided by mean call-option loss for the agent and Black-Scholes.}
    \label{fig:regime_2022_state_and_hedge}
\end{figure}

The market path makes the exposure clear. In the OptionsDX underlying series, the S\&P 500 falls by $-19.9\%$ during 2022. A policy that shorts less underlying earns less hedge profit when calls lose value. Figure \ref{fig:regime_2022_state_and_hedge} measures this channel.

Index-down/call-down intervals account for $54.7\%$ of all 2022 hedge intervals, the largest share in the walk-forward sample. The agent's smaller short-index position recovers only $0.92$ of the mean call loss in these states, whereas Black-Scholes recovers about $1.01$. In P\&L units, the agent averages approximately $-0.045$ per such interval, while Black-Scholes is close to flat at $0.005$.

\begin{figure}[H]
    \centering
    \includegraphics[width=0.62\textwidth]{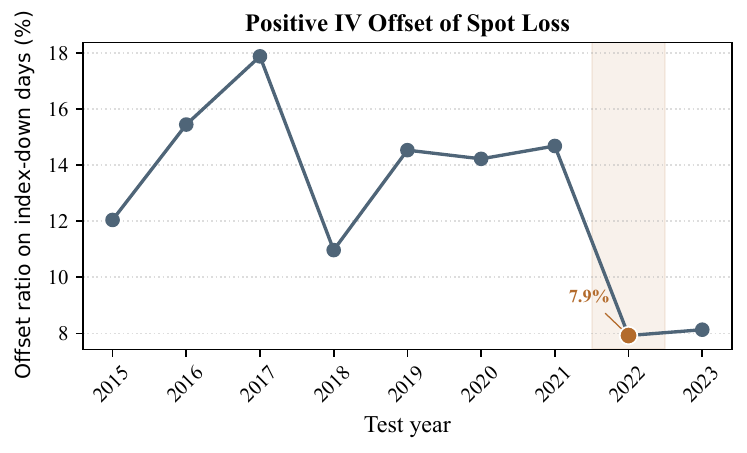}
    \caption{Volatility revaluation in the 2022 reward failure. The figure reports, on index-down hedge intervals, positive implied-volatility revaluation of the call as a fraction of the absolute negative spot component. Higher values mean that implied-volatility increases cushion more of the spot-driven call loss.}
    \label{fig:regime_2022_vega_parachute}
\end{figure}

Figure \ref{fig:regime_2022_vega_parachute} shows why the volatility channel does not rescue the underhedged portfolio. Positive IV revaluations offset only $7.9\%$ of the negative spot revaluation, the lowest value in the pre-2023 sample.

On index-down intervals in 2022, the mean spot component of call revaluation is $-0.520$, while the mean IV component is only $0.014$.  Even when the index falls and IV rises, call P\&L is positive in only $5.5\%$ of intervals, and mean call P\&L remains negative at $-0.555$.

Reward accounting combines the state-frequency and volatility-translation channels. Index-down/call-down intervals contribute $-5.47$ reward units per episode to the agent-minus-Black-Scholes difference; all other states contribute $+3.56$, leaving the observed $-1.90$. The year is therefore a timing failure under the asymmetric objective, not a generic high-variance failure.

\subsection{A Symmetric-Variance Failure in 2023}

The 2023 result presents a different failure mode. Black-Scholes becomes an unusually good symmetric-variance hedge because option P\&L is driven mostly by spot moves and only weakly by the volatility channel. Its short-index hedge therefore offsets the call leg very tightly. The agent continues to underhedge, leaving residual long exposure to the index. That exposure does not hurt downside variance and can help reward or mean P\&L in this particular year, but it increases ordinary variance because ordinary variance also counts favorable outliers as dispersion.

This is why the metric signature looks unusual. The agent does not significantly lose accumulated reward and significantly improves downside variance, with a log downside-variance ratio of $-0.168$. The loss appears only under ordinary terminal variance: the log agent-to-Black-Scholes variance ratio is $0.527$, significant at the 1\% level. The ratio is large because Black-Scholes leaves only $0.049$ of terminal P\&L variance, not because the agent's variance explodes.

\begin{figure}[H]
    \centering
    \includegraphics[width=\textwidth]{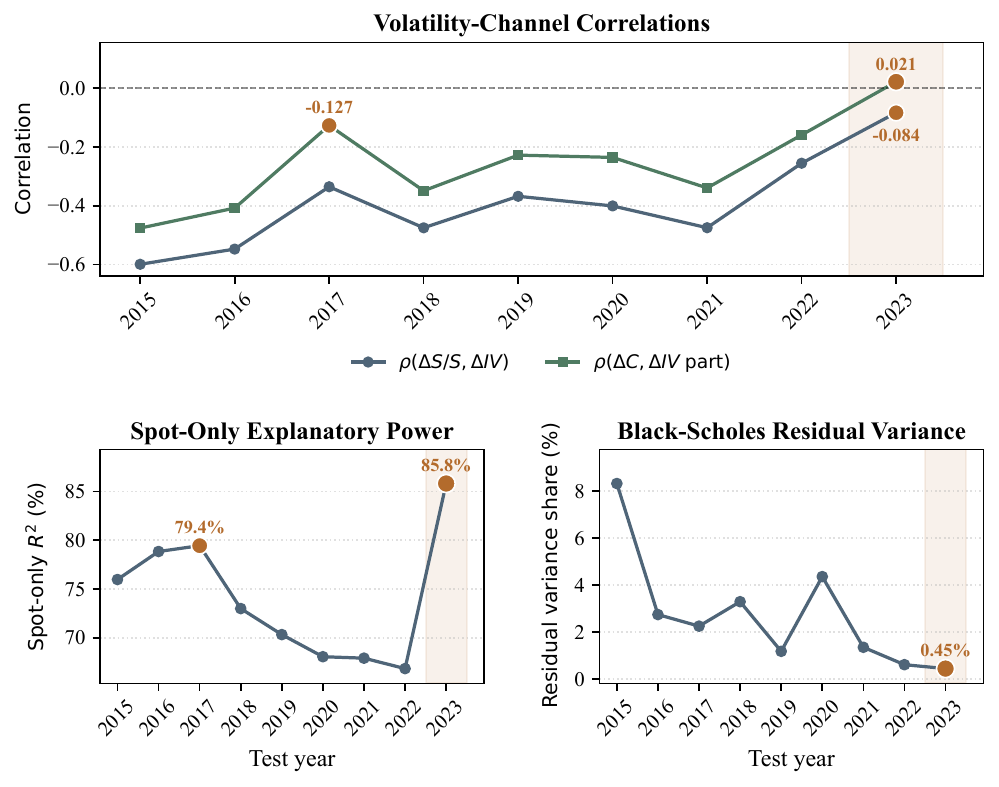}
    \caption{Reduced-form mechanism behind the 2023 ordinary-variance failure. The top panel reports annual correlations for the volatility channel. The bottom-left panel reports the $R^2$ from an annual spot-only regression of daily option P\&L on spot returns. The bottom-right panel reports the Black-Scholes terminal residual variance share after option-leg and hedge-leg covariance cancellation.}
    \label{fig:regime_2023_variance_mechanism}
\end{figure}

Figure \ref{fig:regime_2023_variance_mechanism} gives the evidence in three steps. First, the spot-only regression in the bottom-left panel explains $85.8\%$ of option-price changes, the highest value in the sample. Second, the top panel shows a weak volatility channel: the spot-IV relation is close to zero, with $\rho(\Delta S/S,\Delta\sigma)=-0.084$, and the hedge-relevant relation between call-price changes and the IV revaluation component is also near zero, with $\rho(\Delta C,\Delta IV\text{ part})=0.021$. Third, the bottom-right panel shows the consequence: after option-leg and hedge-leg covariance cancellation, Black-Scholes leaves the smallest residual variance share in the walk-forward sample.

Table \ref{tab:regime_bs_cancellation} reports the same terminal accounting. The 2023 Black-Scholes variance is not low because option prices barely move: option-leg variance is $5.713$ and hedge-leg variance is $5.254$. It is low because the covariance term is $-10.918$, leaving only $0.049$ of terminal P\&L variance, the second smallest in the sample. Black-Scholes leaves only $0.45\%$ of gross option-plus-hedge variance uncancelled.

\begin{table}[H]
    \centering
    \caption{Black-Scholes hedge-account variance decomposition by year. BS variance is the sum of option variance, hedge variance, and $2\,\mathrm{Cov}$. A small BS variance therefore indicates tight cancellation between the option leg and the Black-Scholes hedge leg.}
    \label{tab:regime_bs_cancellation}
    \small
    \input{tables/regime_bs_cancellation.tex}
\end{table}

The agent's variance loss follows from applying its usual correction in this spot-dominated regime. In 2023 the mean Agent--BS hedge gap is $-0.019$, and the agent is underhedged in $67.3\%$ of hedge intervals. On spot-up paths, the call gains value and the Black-Scholes short hedge gives back much of that gain. The underhedged agent gives back less, creating favorable terminal outliers. These outliers are not a problem for the local downside objective and do not prevent the downside-variance improvement, but ordinary variance penalizes them. The 2023 result is therefore a symmetric-variance failure caused by unusually tight Black-Scholes cancellation and the agent's residual long spot exposure.

\subsection{The 2017 Low-Volatility Analogue}

The 2017 result is a milder analogue of 2023. Reward improves by $0.409$ and downside variance falls, but ordinary variance is elevated in all three final-style seeds and significant at 5\% in one of them, see Section \ref{sec:robust_seed}. The mechanism is again weak volatility translation and strong spot dominance: as Figure \ref{fig:regime_2023_variance_mechanism} shows, the IV revaluation correlation is only $-0.127$, and a spot-only regression explains $79.4\%$ of option-price changes.

\begin{figure}[H]
    \centering
    \includegraphics[width=0.62\textwidth]{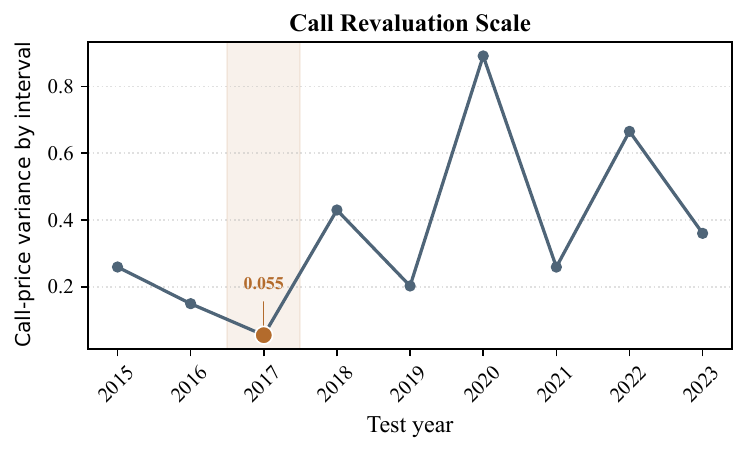}
    \caption{Call revaluation scale in the 2017 ordinary-variance failure. The figure reports, by test year, the variance of daily call-price changes across traded walk-forward hedge intervals without bootstrap resampling.}
    \label{fig:regime_2017_variance_mechanism}
\end{figure}

Figure \ref{fig:regime_2017_variance_mechanism} adds the scale: call-price-change variance is only $0.055$, the lowest in the sample. Table \ref{tab:regime_bs_cancellation} shows that Black-Scholes terminal variance is correspondingly small, $0.043$. The option leg is quiet, and the benchmark has little residual dispersion left for the agent to remove.

The 2017 actor also has a small localized overhedging pocket, but Table \ref{tab:regime_2017_cluster_mechanism} shows that this is not the main source of the variance gap. The largest positive contributors are mostly underhedged clusters; the two largest also have positive agent-minus-Black-Scholes P\&L and reward.

For episode $i$, define the episode-level contribution to the variance gap as
\[
    v_i
    =
    \frac{
        \left(P^{A}_i-\bar P^{A}\right)^2
        -
        \left(P^{BS}_i-\bar P^{BS}\right)^2
    }{N-1},
\]
where $P^{A}_i$ and $P^{BS}_i$ are terminal P\&Ls for the agent and Black-Scholes, and $N$ is the number of 2017 test episodes. For a 21-day start-date cluster $g$, let $V_g=\sum_{i\in g}v_i$. The last column of Table \ref{tab:regime_2017_cluster_mechanism} reports $V_g$ as a share of the sum of positive cluster contributions. The two largest positive contributors have negative mean Agent--BS delta gaps and positive average agent-minus-Black-Scholes terminal P\&L and reward, so they are not overhedged clusters.

\begin{table}[H]
    \centering
    \caption{Largest positive cluster contributions to the 2017 ordinary-variance gap. A cluster is a common 21-day episode start date. $\Delta$PnL and $\Delta$Reward are average agent-minus-Black-Scholes terminal P\&L and accumulated reward within the cluster. Excess Var. is $V_g/\sum_{h:V_h>0}V_h$, where $V_g$ is the cluster contribution defined in the text. Negative Agent--BS $\Delta$ means that the agent shorts less underlying than Black-Scholes on average within the cluster.}
    \label{tab:regime_2017_cluster_mechanism}
    \small
    \input{tables/regime_2017_cluster_mechanism.tex}
\end{table}

The three cases therefore have one lesson: the learned correction is interpretable, but its value depends on the realized option-surface regime and on the metric used to judge the hedge.

\section{Symbolic Policy Distillation}
\label{sec:symbolic_distillation}

This section asks whether the learned correction can be compressed into an explicit traded formula. Symbolic regression often preserves, and in some metrics strengthens, the Black-Scholes comparison, but it does not produce a uniformly superior replacement for the neural policy.

\subsection{Symbolic Regression and Distillation Procedure}

Symbolic regression searches over analytical expressions and balances fit against expression complexity. The inputs are forward moneyness $m$, time to maturity $\tau$, and implied volatility $\sigma$. The symbolic policy is fit as a correction to Black-Scholes delta:
\begin{equation}
    \widehat{\Delta}^{\,SR}(m,\tau,\sigma)
    =
    \mathrm{clip}\left(
        \Delta^{BS}(m,\tau,\sigma;r,q)
        +
        g(m,\tau,\sigma),
        0,1
    \right),
    \label{eq:sr_delta_residual}
\end{equation}
where $g$ is the symbolic residual. This anchors the formula to the Black-Scholes geometry and asks symbolic regression to learn only the state-dependent correction.

The distillation experiment is walk-forward. For each test year, the trained actor is frozen and queried only on training-support states and a pre-specified off-path probe domain. Validation and test years are not used to fit formulas.

Three formula families are fit for each yearly agent: raw uniform, smooth uniform, and smooth focus. The first fits the raw actor residual; the latter two fit a kernel-smoothed residual, with the smooth-focus family changing the sampling weights as described in Appendix \ref{sec:appendix_focus_sampling}. The smoothed policy is
\begin{equation}
    \Delta^{S}(m,\tau,\sigma)
    =
    \mathrm{clip}\left(
        \Delta^{BS}(m,\tau,\sigma;r,q)
        +
        \widehat{s}(m,\tau,\sigma),
        0,1
    \right),
    \label{eq:smoothed_policy}
\end{equation}
where $\widehat{s}$ is the kernel-smoothed residual. The smoothed policy itself is also traded, so the analysis distinguishes the raw agent, smoothed agent, and symbolic formula.

Symbolic regression produces a Hall of Fame of expressions for each family and year. The main selection rule is parsimonious: among pooled formulas whose validation MAE against the traded raw agent is within 10\% of the best pooled validation MAE, choose the lowest-complexity expression. The chosen formula is then evaluated against the raw agent, smoothed agent, and Black-Scholes. The test year is not used for fitting or selection. Section \ref{sec:robust_symbolic_selection} checks alternative validation rules.

\subsection{The Smoothed Agent as a Diagnostic Object}

Smoothing stays close to the raw actor pointwise: average delta MAE is 0.006 and correlations exceed 0.997 in every year. Table \ref{tab:distill_smoother_vs_agent_yearly} shows that this small change can still matter in trading, especially for reward.

\begin{table}[H]
    \centering
    \caption{Smoothed policy relative to the raw traded agent. Entries are year-by-year smoothed-minus-raw-agent results; $^{*}$, $^{**}$, and $^{***}$ denote 10\%, 5\%, and 1\% two-sided two-stage bootstrap significance, respectively. Reward, CVaR, and mean P\&L entries are differences, so positive values are favorable; variance entries are log ratios, so negative values are favorable.}
    \label{tab:distill_smoother_vs_agent_yearly}
    \input{tables/distill_smoother_vs_agent_yearly.tex}
\end{table}

Table \ref{tab:distill_smoother_vs_bs_summary} shows that smoothing preserves the main Black-Scholes comparison and increases significant CVaR wins from one to four.

\begin{table}[H]
    \centering
    \caption{Raw and smoothed policies relative to Black-Scholes. Entries are averages across the nine test years. Parentheses report favorable/unfavorable years significant at the 5\% two-sided two-stage bootstrap level. Reward and CVaR entries are policy-minus-Black-Scholes differences; variance entries are log policy-to-Black-Scholes ratios, so negative values are favorable.}
    \label{tab:distill_smoother_vs_bs_summary}
    \input{tables/distill_smoother_vs_bs_summary.tex}
\end{table}

The smoothed policy is therefore a useful diagnostic object that has some regularization properties, but not a uniformly better hedge.

\subsection{Fit and Trading Performance}

The final rule selects one formula per test year. Table \ref{tab:distill_parsimonious_fit_summary} reports the selected family, complexity, and pointwise fit. The pooled 10\% rule chooses smooth-uniform formulas in four years, raw-uniform formulas in three, and smooth-focus formulas in two. Average selected complexity is 9.6. The rule often rejects the absolute best-fitting expression for a much shorter one; in 2015, for example, complexity falls from 24 to 6 while remaining inside the validation-error band.

\begin{table}[H]
    \centering
    \caption{Selected symbolic formula family by test year. Complexity is the expression-tree complexity reported by symbolic regression. Validation MAE is measured on the validation-year states against the actually traded raw agent. Test MAE, the 95\% absolute error, and the test correlation are measured on the test-year states against the same raw agent action.}
    \label{tab:distill_parsimonious_fit_summary}
    \input{tables/distill_parsimonious_fit_summary.tex}
\end{table}

The selected formulas fit the raw actor closely but not perfectly: average test MAE is 0.021 and average test correlation is 0.995.

Table \ref{tab:distill_parsimonious_trading_summary} shows that the formulas remain economically meaningful when traded. Relative to Black-Scholes, they improve average reward, downside variance, and CVaR, while ordinary variance is neutral on average. The selected formulas also have an important non-deterioration property relative to agent-vs-Black-Scholes comparisons. Across all metrics the parsimonious rule never creates a statistically significant formula-versus-Black-Scholes loss in a year and metric where the agent did not already suffer a significant loss. It also never rescues such a loss: the formula loses on reward in 2022, as the agent does, and loses on ordinary variance in 2023, as the agent does.

\begin{table}[H]
    \centering
    \caption{Trading performance of the selected symbolic formulas and the raw agent. Entries are averages across the nine test years. Parentheses report favorable/unfavorable years significant at the 5\% two-sided two-stage bootstrap level. Reward and CVaR entries are left-minus-right differences; variance entries are log left-to-right ratios, so negative values are favorable.}
    \label{tab:distill_parsimonious_trading_summary}
    \input{tables/distill_parsimonious_trading_summary.tex}
\end{table}

\begin{figure}[H]
    \centering
    \includegraphics[width=\textwidth]{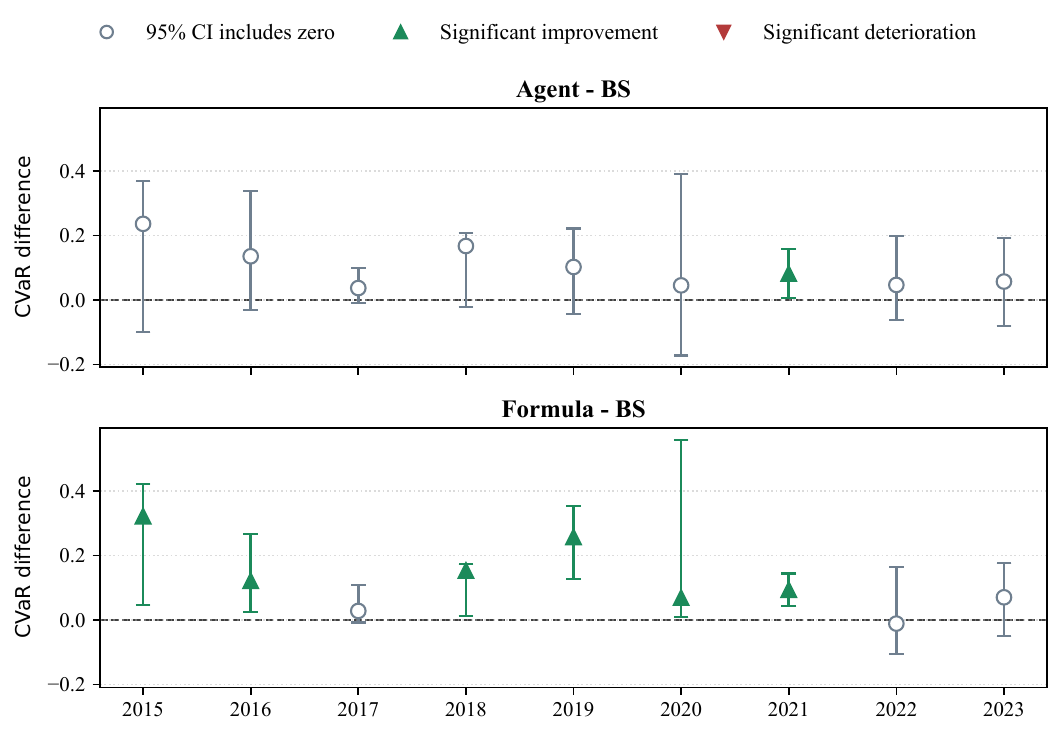}
    \caption{CVaR comparison with Black-Scholes for the raw agent and the selected symbolic formula. Points show policy-minus-Black-Scholes 5\% CVaR differences; bars are two-stage bootstrap 95\% confidence intervals. Positive values are favorable.}
    \label{fig:distill_agent_formula_cvar_vs_bs_pair}
\end{figure}

Figure \ref{fig:distill_agent_formula_cvar_vs_bs_pair} shows the strongest symbolic result: formula CVaR is favorable in eight of nine years and significant in six. 

The improvement is related to smoothing but not reducible to it: four significant formula years come from smoothed-target families and two from the raw family.

\begin{figure}[H]
    \centering
    \includegraphics[width=\textwidth]{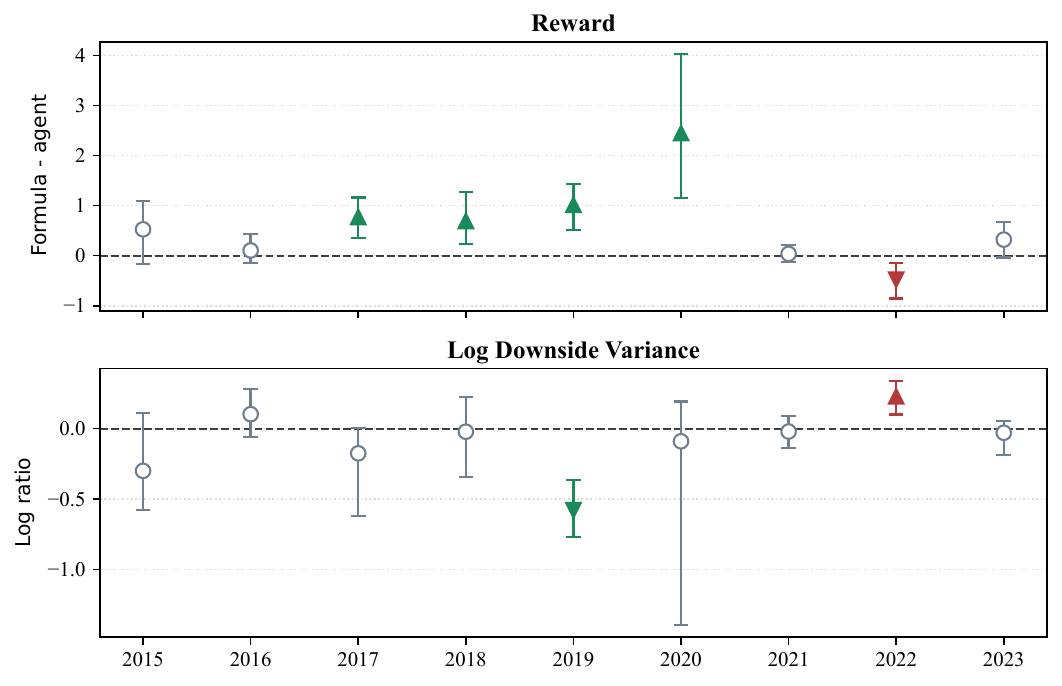}
    \caption{Trading performance of the selected symbolic formula relative to the neural agent. Points show formula-minus-agent performance; bars are two-stage bootstrap 95\% confidence intervals. Positive values are favorable for reward; negative values are favorable for log downside variance.}
    \label{fig:distill_parsimonious_smooth_focus_vs_agent_pair}
\end{figure}

Relative to the neural agent itself, the evidence is mixed. Figure \ref{fig:distill_parsimonious_smooth_focus_vs_agent_pair} shows that formulas can improve reward relative to the actor in 2017--2020, but they lose to the actor in 2022. Symbolic distillation regularizes the hedge; it does not remove regime fragility\footnote{Appendix Section \ref{sec:appendix_regime_switching} shows that a targeted 2022 switching check does not remove the failure.} and can worsen it in specific regimes, such as 2022.

The distillation result is therefore useful but limited: compact formulas preserve much of the economic signal, but not all of the neural policy's local state dependence.

\subsection{Structure of the Selected Corrections}

The selected expressions differ by year, but share a simple structure: they lower Black-Scholes delta over most of the state space, with the size of the reduction varying mainly with moneyness and implied volatility. On each formula's distillation support, the symbolic policy lies below Black-Scholes in 94.2\% of states on average. The mean correction is $-0.055$, with median $-0.048$.

This form is consistent with the standard minimum-variance delta intuition under stochastic volatility. In a one-period linearization,
\[
    dC \approx \Delta^{BS} dS + \nu d\sigma,
\]
where $\nu$ is vega. The stock hedge that minimizes the variance of the residual option-price change is therefore
\[
    h^{MV}
    =
    \Delta^{BS}
    +
    \nu \frac{\mathrm{Cov}(d\sigma,dS)}{\mathrm{Var}(dS)} .
\]
For equity-index options, spot and implied volatility are typically negatively correlated. With positive call vega, this covariance term pushes the minimum-variance hedge below the practitioner Black-Scholes delta. This is the mechanism emphasized by \citet{hull2017optimal}; related smile-adjusted delta work also points to simple, state-dependent corrections \citep{alexander2012regime}.

The symbolic formulas recover a reduced-form version of this idea. A particularly transparent selected expression is the 2016 smooth-uniform formula,
\begin{equation}
    \widehat{\Delta}^{\,SR}_{2016}
    =
    \mathrm{clip}\left(
        \Delta^{BS}
        +
        \sigma\left(e^{m}-2.949\right),
        0,1
    \right).
    \label{eq:sr_2016_example}
\end{equation}
This expression is a useful representative example because it has low complexity, strong out-of-sample performance against Black-Scholes, and a direct moneyness-volatility structure. Around $m=1$,
\[
    \sigma(e^m-2.949)
    \approx
    \sigma(2.718(m-1)-0.231)
    =
    2.718 \cdot \sigma(m - 1.085).
\]
Near at-the-money states the residual is negative, so the symbolic policy hedges less than Black-Scholes. The correction becomes less negative as forward moneyness rises and scales with implied volatility.

The simplest rules are almost pure moneyness shifts, for example $m-1.045$ in 2017 and $m-1.063$ in 2019. Other years add a volatility scale, such as \eqref{eq:sr_2016_example} or $\sqrt{\sigma}$ terms in 2020 and 2023. Across formulas, the correction is positively associated with forward moneyness and negatively associated with implied volatility: average correlations are 0.565 with $m$, $-0.346$ with $\sigma$, and only $-0.062$ with maturity. The dominant symbolic pattern is a moneyness-centered delta haircut, stronger in higher-volatility regions in several years.

This interpretation has limits. The formulas imitate a downside-reward actor, not a classical minimum-variance delta, see Section \ref{sec:robust_hull_white}. They combine an economically familiar spot-volatility correction with regularization of the neural policy.

\section{Robustness Checks}

\subsection{Random-Seed Robustness}
\label{sec:robust_seed}

The main walk-forward results use one final training run per test year. Appendix Tables \ref{tab:robust_seed_new_seed} and \ref{tab:robust_seed_1_seed} report two additional final-style seed runs. The qualitative conclusions are stable. Downside-variance ratios are negative in almost every year and seed, with only one exception in 2022. Ordinary-variance failures persist: 2023 is positive in all three runs, and 2017 is elevated in all three runs and significant at 5\% in one run. Reward is less stable, but the key failure remains: 2022 is negative and significant in every run.

\subsection{Long-Horizon Policy Stress Test}
\label{sec:robust_long_horizon}

\begin{figure}[H]
    \centering
    \includegraphics[width=\textwidth]{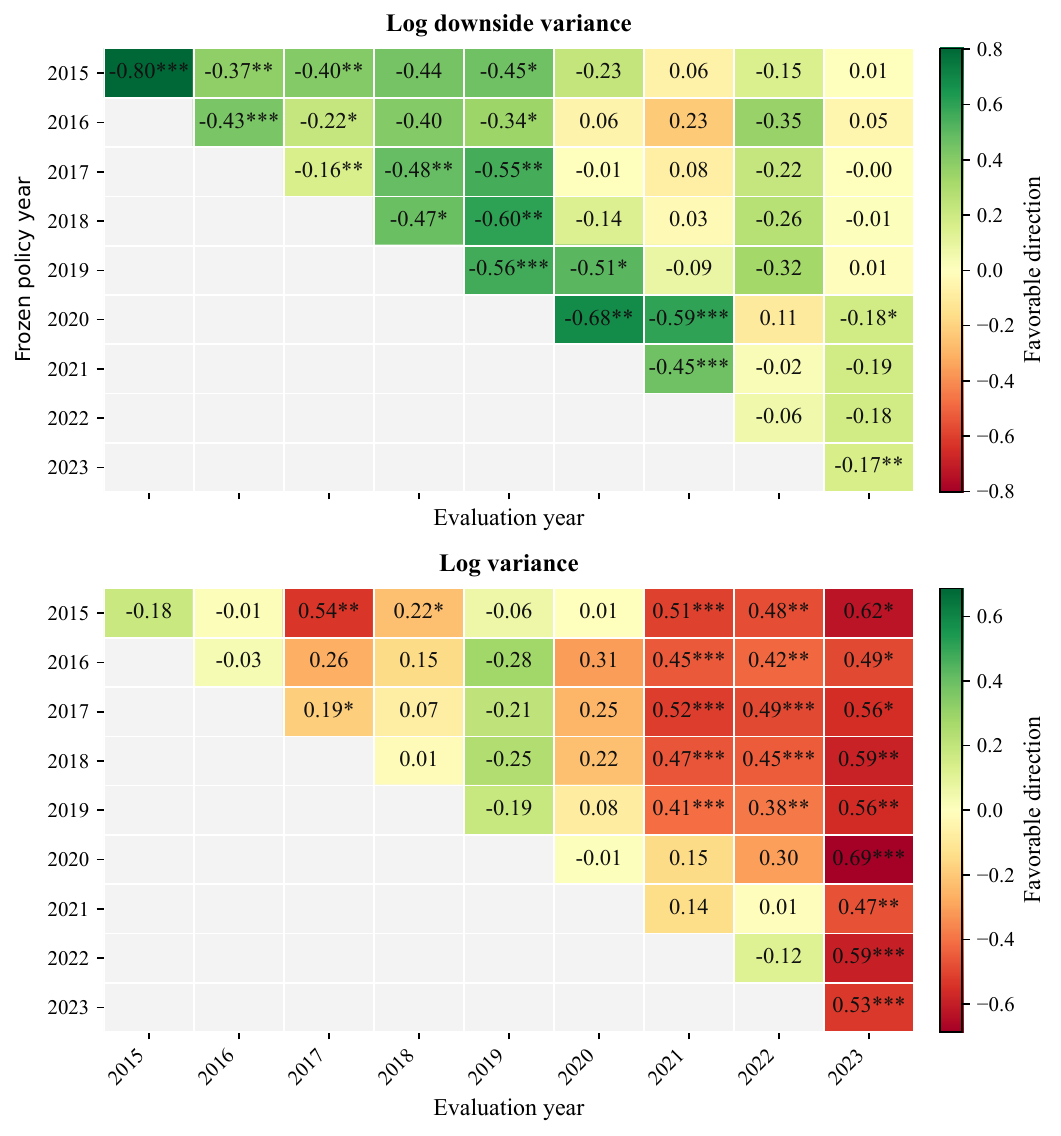}
    \caption{Long-horizon policy stress test for downside and ordinary variance. Rows identify the frozen policy by its original walk-forward test year; columns identify the later evaluation year. Cells report agent performance relative to Black-Scholes. For both panels the entries are log agent-to-Black-Scholes ratios, so negative values are favorable. Green cells are favorable for the agent, red cells are unfavorable, and $^{*}$, $^{**}$, and $^{***}$ denote 10\%, 5\%, and 1\% two-sided two-stage bootstrap significance, respectively.}
    \label{fig:robust_long_horizon_variance}
\end{figure}

The long-horizon stress test asks how a frozen policy behaves when carried into later regimes. For each policy whose original test year is $Y$, the checkpoint is re-evaluated on $Y,Y+1,\ldots,2023$ without retraining or changing the selection rule. Each panel in Figure \ref{fig:robust_long_horizon_variance} is triangular, so later years receive more weight; the exercise is a stress test, not a replacement for the walk-forward evidence.

Figure \ref{fig:robust_long_horizon_variance} shows that downside variance is more persistent than ordinary variance. Across 45 frozen-policy/evaluation-year pairs, the downside-variance ratio is favorable in 36 cases and significantly favorable in 13, with no significant unfavorable case. Ordinary variance is favorable in only 10 cases and significantly unfavorable in 17. Appendix Figure \ref{fig:robust_long_horizon_reward_mean} shows that mean terminal P\&L remains favorable in 43 of 45 pairs, while reward is much less stable. The robust component is therefore a downside-oriented correction that can survive regime drift, while reward and ordinary variance are more sensitive to the evaluation regime.

\subsection{Symbolic Selection Rule Robustness}
\label{sec:robust_symbolic_selection}

Table \ref{tab:robust_selection_rule_summary} compares the parsimonious 10\% fit rule with four validation-only alternatives: best validation fit, best validation reward, best validation CVaR, and best validation downside variance. In every case, the test year is untouched until the selected formula is evaluated.

\begin{table}[H]
    \centering
    \caption{Robustness of symbolic policy performance to the validation selection rule. ``Comp.'' is average expression-tree complexity and MAE is average test mean absolute delta error against the traded agent. Trading columns report averages across test years; parentheses give favorable/unfavorable years significant at the 5\% two-sided two-stage bootstrap level. Reward entries are left-minus-right differences. DownVar entries are log left-to-right downside-variance ratios, so negative values are favorable.}
    \label{tab:robust_selection_rule_summary}
    \small
    \input{tables/robust_selection_rule_summary.tex}
\end{table}

The conclusion is not an artifact of the parsimonious rule. All five rules improve reward relative to Black-Scholes on average, with five or six significant favorable years and one unfavorable year. All five also produce negative average downside-variance ratios, with seven or eight significant favorable years and no unfavorable years. Best validation fit lowers test MAE only from $0.021$ to $0.020$ while raising average complexity from $9.6$ to $16.6$, so extra complexity buys little. Relative to the neural agent, all rules remain mixed, confirming that symbolic formulas are compact regularized hedges, not uniformly superior replacements.

\subsection{Hull-White-Style Delta-Correction Benchmark}
\label{sec:robust_hull_white}

A natural concern is that Black-Scholes is too weak a benchmark. If the learned hedge mainly corrects for the interaction between spot moves and volatility changes, a classical minimum-variance delta correction should capture a material part of the effect. We therefore add a Hull-White-style benchmark, following the correction logic of \citet{hull2017optimal}. For each test year, the correction is estimated on the immediately preceding validation year and then frozen out of sample:
\[
\Delta^{HW}_t =
\operatorname{clip}\left(
\Delta^{BS}_t
+ \frac{\nu_t}{S_t\sqrt{\tau_t}}
\left(a+b\Delta^{BS}_t+c(\Delta^{BS}_t)^2\right),
0,1
\right),
\]
where $\nu_t$ is Black-Scholes vega and the polynomial coefficients are fit from validation residuals. This benchmark is deliberately simple, but it is a meaningful comparator because it gives the parametric correction access to the most recent pre-test spot--volatility relation.\footnote{Appendix Table \ref{tab:robust_haircut_summary} reports a validation-selected scalar haircut, $\lambda\Delta^{BS}$.}

Appendix Tables \ref{tab:robust_hull_white_vs_bs}--\ref{tab:robust_hull_white_vs_formula} give the full year-by-year comparisons. The Hull-White-style correction improves on Black-Scholes in several years, especially for downside variance, but it does not reproduce the asymmetric hedging profile of either the neural policy or the selected symbolic formulas. Table \ref{tab:robust_hull_white_summary} summarizes the direct comparison. On average, the benchmark is worse than the agent on CVaR and downside variance, and it is also worse than the selected formulas on downside variance. The result is informative: a low-dimensional volatility correction explains part of the direction of the learned policy, but not the economic performance generated by the asymmetric objective.

\begin{table}[H]
    \centering
    \caption{Hull-White-style benchmark relative to the learned hedges. Entries are averages across test years. Reward and CVaR are left-minus-right differences; log downside variance and log variance are log ratios. Negative values are favorable for variance metrics. Parentheses report the number of favorable/unfavorable years significant at the 5\% two-sided two-stage bootstrap level.}
    \label{tab:robust_hull_white_summary}
    \small
    \input{tables/robust_hull_white_summary.tex}
\end{table}

\section{Conclusion}

This paper contributes to empirical deep hedging by moving from performance comparison to mechanism analysis. Under a local downside-shortfall reward, the TD3 agent learns a clear economic correction: it usually hedges long SPX calls by shorting less underlying than a daily-updated Black-Scholes delta. The direction is consistent with the equity-index volatility channel: when index declines are partly cushioned by higher implied volatility, a smaller short hedge can be preferable for a downside-oriented objective.

The correction is useful, but conditional. In the walk-forward tests it often improves accumulated reward, downside variance, or tail-risk metrics, yet it is not a universal variance-reduction rule. In 2022, adverse daily states punish the underhedge and the volatility channel does not compensate enough for call losses. In 2023, option P\&L is unusually spot-dominated and Black-Scholes cancellation is unusually tight, so favorable residual exposure becomes costly under ordinary variance. These failures identify the regimes and metrics for which the learned correction stops helping.

Symbolic regression compresses much of the neural policy into compact closed-form formulas that can be traded out of sample. The selected formulas preserve, and in some metrics strengthen, the Black-Scholes comparison, especially for reward, downside variance, and CVaR, but they inherit fragility in difficult regimes. The main lesson is therefore methodological as well as empirical: deep hedging under asymmetric objectives should be evaluated jointly through the optimized reward, terminal risk metrics, regime diagnostics, and policy interpretation. Future work should compare these policies with stronger model-based hedges, other option types, downside thresholds, and transaction-cost regimes.

\clearpage
\bibliographystyle{plainnat}
\bibliography{Bibliography}

\clearpage
\appendix

\section{Symbolic-Distillation Sampling Details}
\label{sec:appendix_focus_sampling}

The smooth-focus family uses the same smoothed target as the smooth-uniform family, but changes the distribution of symbolic-regression probe points. It combines two standard devices: stratified sampling over state-space cells and importance-weighted sampling toward regions expected to matter more for fitting the hedge correction \citep{cochran1977sampling,owen2013monte}.

Let $m$ denote forward moneyness, $\sigma$ implied volatility, and $\tau$ time to maturity in years. For each candidate point in the training/probe distillation pool, define the absolute hedge-correction gap
\[
    g_i = \left|\Delta_i^{S}-\Delta_i^{BS}\right|,
\]
where $\Delta_i^{S}$ is the smoothed-agent delta and $\Delta_i^{BS}$ is the Black-Scholes delta. Define also a gamma-like state proxy
\[
    \Gamma_i
    =
    \frac{
        \exp\left[-\frac{1}{2}\left\{\frac{\log(m_i)}{\sigma_i\sqrt{\tau_i}}\right\}^{2}\right]
    }{
        \sigma_i\sqrt{\tau_i}
    },
\]
with the denominator clipped away from zero in the implementation. Both $g_i$ and $\Gamma_i$ are normalized by dividing by their respective 95th percentiles and clipping the resulting values to $[0,3]$. Denote the normalized quantities by $\tilde g_i$ and $\tilde \Gamma_i$.

The fixed focus indicator is
\[
    I_{\mathcal F,i}
    =
    \mathbf{1}\{m_i\in[1.03,1.15],\ \sigma_i\in[0.20,0.38],\ T_i\in[25,90]\},
\]
where $T_i$ is maturity in calendar days. The smooth-focus sampling weight is then
\[
    w_i
    =
    1 + 4\tilde g_i + 1.5\tilde\Gamma_i + 4 I_{\mathcal F,i}.
\]
The final sample consists of two parts. First, 60\% of the rows are drawn by weighted stratified sampling across cells in forward moneyness, implied volatility, and maturity. The bin edges are
\[
\begin{aligned}
    m &: 0.70,\,0.85,\,0.95,\,1.00,\,1.03,\,1.07,\,1.12,\,1.20,\,1.40,\\
    \sigma &: 0.00,\,0.10,\,0.12,\,0.16,\,0.20,\,0.24,\,0.28,\,0.35,\,0.50,\,0.85,\\
    T &: 0,\,14,\,21,\,35,\,50,\,65,\,80,\,100,\,140\ \text{days}.
\end{aligned}
\]
Second, the remaining 40\% of the rows are drawn from the residual pool with probabilities proportional to $w_i$. The rule is fixed before validation and test evaluation and uses only the training/probe distillation pool.

\section{Additional Robustness Tables}
\captionsetup[table]{skip=2pt}

\subsection{Random-Seed Robustness Tables}
\label{sec:appendix_seed_tables}

Tables \ref{tab:robust_seed_new_seed} and \ref{tab:robust_seed_1_seed} report the two additional final-style seed runs discussed in Section \ref{sec:robust_seed}; the format follows Table \ref{tab:wf_metric_summary}. Reward, CVaR, and mean P\&L are agent-minus-Black-Scholes differences; the variance columns are log agent-to-Black-Scholes ratios. $^{*}$, $^{**}$, and $^{***}$ denote 10\%, 5\%, and 1\% two-sided two-stage bootstrap significance, respectively.

\begin{table}[H]
    \centering
    \caption{Random-seed robustness: second final-style seed.}
    \label{tab:robust_seed_new_seed}
    \small
    \input{tables/robust_seed_metrics_new_seed_final_WF_exp1_k1_test.tex}
\end{table}

\begin{table}[H]
    \centering
    \caption{Random-seed robustness: third final-style seed.}
    \label{tab:robust_seed_1_seed}
    \small
    \input{tables/robust_seed_metrics_1_seed_final_WF_exp1_k1_test.tex}
\end{table}

\subsection{Long-Horizon Reward and Mean P\&L Panels}
\label{sec:appendix_long_horizon}

\begin{figure}[H]
    \centering
    \includegraphics[width=\textwidth]{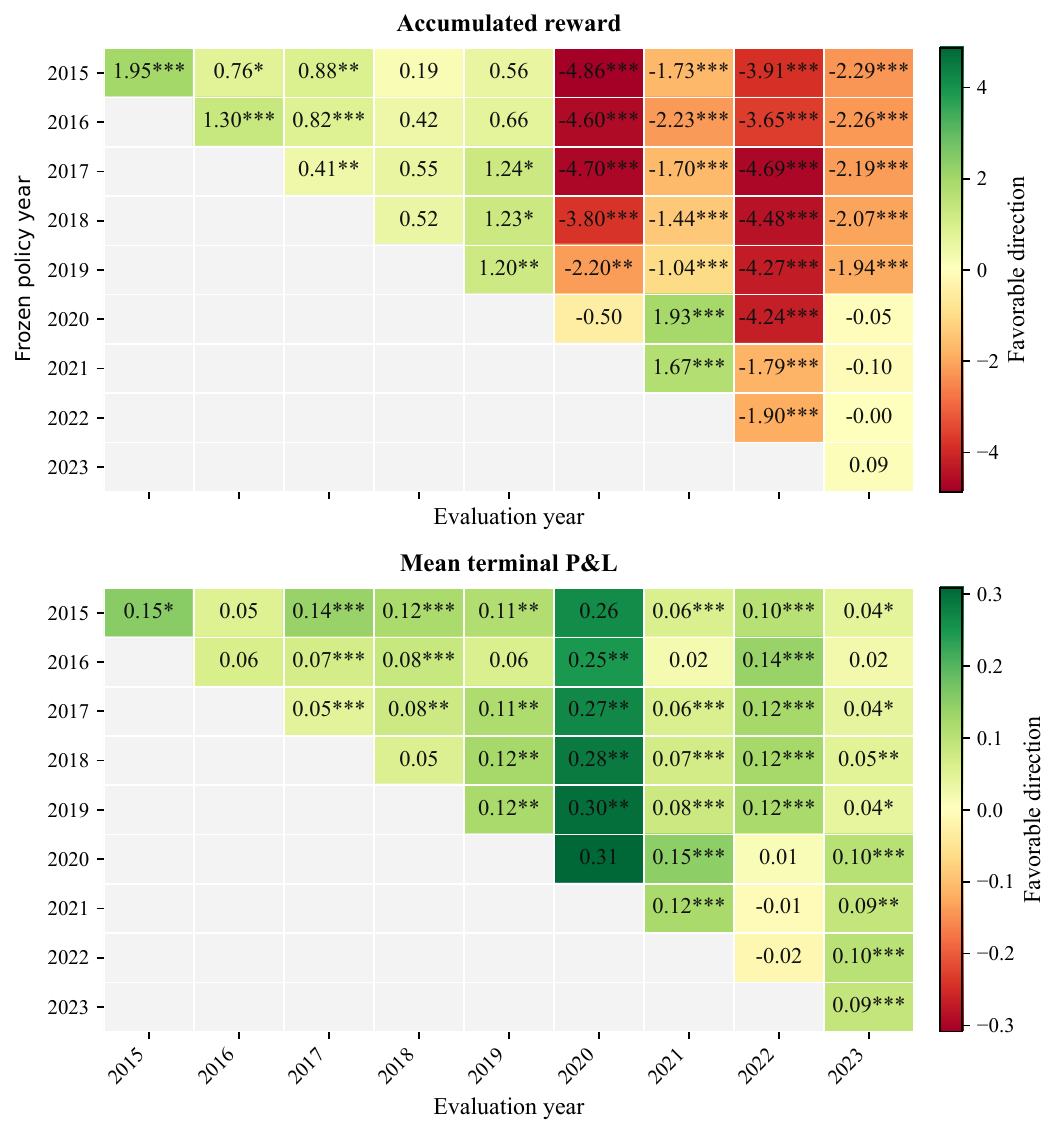}
    \caption{Long-horizon policy stress test for accumulated reward and mean terminal P\&L. Cells report agent-minus-Black-Scholes point estimates. Green cells are favorable for the agent, red cells are unfavorable, and $^{*}$, $^{**}$, and $^{***}$ denote 10\%, 5\%, and 1\% two-sided two-stage bootstrap significance, respectively.}
    \label{fig:robust_long_horizon_reward_mean}
\end{figure}

Figure \ref{fig:robust_long_horizon_reward_mean} reports the long-horizon reward and mean terminal P\&L panels discussed in Section \ref{sec:robust_long_horizon}. The triangular structure is the same as in Figure \ref{fig:robust_long_horizon_variance}: rows are frozen policy years and columns are later evaluation years.

\subsection{Hull-White-Style Delta-Correction Benchmark}
\label{sec:appendix_hull_white}

Tables \ref{tab:robust_hull_white_vs_bs}--\ref{tab:robust_hull_white_vs_formula} report the full yearly comparisons discussed in Section \ref{sec:robust_hull_white}. The format follows Table \ref{tab:wf_metric_summary}: reward, CVaR, and mean terminal P\&L are differences, variance metrics are log ratios, and $^{*}$, $^{**}$, and $^{***}$ denote 10\%, 5\%, and 1\% two-sided two-stage bootstrap significance, respectively.

\begin{table}[H]
    \centering
    \caption{Hull-White-style benchmark relative to Black-Scholes, including mean P\&L.}
    \label{tab:robust_hull_white_vs_bs}
    \small
    \input{tables/robust_hull_white_vs_bs.tex}
\end{table}

\begin{table}[H]
    \centering
    \caption{Hull-White-style benchmark relative to the neural agent, including mean P\&L.}
    \label{tab:robust_hull_white_vs_agent}
    \small
    \input{tables/robust_hull_white_vs_agent.tex}
\end{table}

\begin{table}[H]
    \centering
    \caption{Hull-White-style benchmark relative to the selected symbolic formula, including mean P\&L.}
    \label{tab:robust_hull_white_vs_formula}
    \small
    \input{tables/robust_hull_white_vs_formula.tex}
\end{table}

\subsection{Scalar Delta-Haircut Benchmark}
\label{sec:appendix_haircut}

The scalar haircut benchmark chooses $\lambda\in\{0.70,0.75,\ldots,1.05\}$ on the validation year and hedges with $\lambda\Delta^{BS}$ in the test year. Table \ref{tab:robust_haircut_summary} shows average reward gains relative to Black-Scholes, but significantly weaker downside performance than the learned hedges.

\begin{table}[H]
    \centering
    \caption{Scalar delta-haircut benchmark. Entries follow the format of Table \ref{tab:robust_hull_white_summary}.}
    \label{tab:robust_haircut_summary}
    \small
    \input{tables/robust_haircut_summary.tex}
\end{table}

\subsection{Regime-Switching Distillation Check}
\label{sec:appendix_regime_switching}

As a targeted diagnostic for 2022, the analysis fits regime-switching symbolic policies with two or three forward-moneyness bands. Table \ref{tab:robust_switching_negative} reports the results. The three-band formula improves test MAE slightly, from $0.012$ to $0.011$, but still loses to the neural agent on downside variance and to Black-Scholes on reward. The 2022 failure is therefore not removed by a simple moneyness partition.

\begin{table}[H]
    \centering
    \caption{Moneyness regime-switching check for 2022. A two-star entry denotes 5\% two-sided two-stage bootstrap significance.}
    \label{tab:robust_switching_negative}
    \resizebox{0.78\textwidth}{!}{\input{tables/robust_switching_negative_result.tex}}
\end{table}

\end{document}

%% file: tables/wf_metric_summary.tex
\begingroup
\renewcommand{\arraystretch}{1.18}
\setlength{\tabcolsep}{5pt}
\begin{tabular}{@{}cccccc@{}}
\toprule
Year & Reward & CVaR 5\% & Mean P\&L & Log Downside Variance & Log Variance \\
\midrule
2015 & $1.946$\sym{***} & $0.236$ & $0.150$\sym{*} & $\mathllap{-}0.802$\sym{***} & $\mathllap{-}0.177$ \\
2016 & $1.302$\sym{***} & $0.136$\sym{*} & $0.061$ & $\mathllap{-}0.429$\sym{***} & $\mathllap{-}0.031$ \\
2017 & $0.409$\sym{**} & $0.037$\sym{*} & $0.046$\sym{***} & $\mathllap{-}0.162$\sym{**} & $0.187$ \\
2018 & $0.523$ & $0.167$ & $0.054$ & $\mathllap{-}0.467$\sym{*} & $0.014$ \\
2019 & $1.205$\sym{**} & $0.102$ & $0.119$\sym{**} & $\mathllap{-}0.557$\sym{***} & $\mathllap{-}0.186$ \\
2020 & $\mathllap{-}0.498$ & $0.045$ & $0.308$ & $\mathllap{-}0.677$\sym{**} & $\mathllap{-}0.009$ \\
2021 & $1.674$\sym{***} & $0.082$\sym{**} & $0.120$\sym{***} & $\mathllap{-}0.451$\sym{***} & $0.142$ \\
2022 & $\mathllap{-}1.905$\sym{**} & $0.047$ & $\mathllap{-}0.016$ & $\mathllap{-}0.062$ & $\mathllap{-}0.117$ \\
2023 & $0.095$ & $0.057$ & $0.087$\sym{***} & $\mathllap{-}0.168$\sym{**} & $0.527$\sym{***} \\
\bottomrule
\end{tabular}
\endgroup

%% file: tables/learned_delta_summary.tex
\begingroup
\renewcommand{\arraystretch}{1.16}
\setlength{\tabcolsep}{7pt}
\begin{tabular}{@{}ccccc@{}}
\toprule
Year & Agent Delta & BS Delta & Agent--BS & Underhedged Share \\
\midrule
2015 & $0.455$ & $0.492$ & $\mathllap{-}0.036$ & $86.2$\% \\
2016 & $0.455$ & $0.490$ & $\mathllap{-}0.035$ & $95.8$\% \\
2017 & $0.521$ & $0.530$ & $\mathllap{-}0.009$ & $57.1$\% \\
2018 & $0.534$ & $0.560$ & $\mathllap{-}0.026$ & $75.4$\% \\
2019 & $0.533$ & $0.560$ & $\mathllap{-}0.027$ & $75.7$\% \\
2020 & $0.474$ & $0.526$ & $\mathllap{-}0.052$ & $82.1$\% \\
2021 & $0.498$ & $0.531$ & $\mathllap{-}0.033$ & $88.1$\% \\
2022 & $0.441$ & $0.486$ & $\mathllap{-}0.045$ & $93.0$\% \\
2023 & $0.677$ & $0.696$ & $\mathllap{-}0.019$ & $67.3$\% \\
\bottomrule
\end{tabular}
\endgroup

%% file: tables/negative_pnl_ranked_summary.tex
\begingroup
\renewcommand{\arraystretch}{1.14}
\setlength{\tabcolsep}{5.5pt}
\begin{tabular}{@{}ccccc@{}}
\toprule
Rank & Agent Loss State & Share & BS Loss State & Share \\
\midrule
1 & S Down / Opt. Down / IV Down & $40.9$\% & S Up / Opt. Up / IV Down & $42.1$\% \\
2 & S Up / Opt. Up / IV Down & $27.8$\% & S Down / Opt. Down / IV Down & $32.2$\% \\
\bottomrule
\end{tabular}
\endgroup

%% file: tables/regime_bs_cancellation.tex
\begingroup
\renewcommand{\arraystretch}{1.12}
\setlength{\tabcolsep}{6.0pt}
\begin{tabular}{@{}ccccc@{}}
\toprule
Year & Option Var. & Hedge Var. & $2\,\mathrm{Cov}$ & BS Var. \\
\midrule
2015 & $3.173$ & $6.022$ & $\mathllap{-}8.430$ & $0.765$ \\
2016 & $2.459$ & $2.869$ & $\mathllap{-}5.182$ & $0.146$ \\
2017 & $1.005$ & $0.892$ & $\mathllap{-}1.855$ & $0.043$ \\
2018 & $6.578$ & $8.149$ & $\mathllap{-}14.242$ & $0.485$ \\
2019 & $4.500$ & $5.194$ & $\mathllap{-}9.579$ & $0.115$ \\
2020 & $15.814$ & $23.982$ & $\mathllap{-}38.061$ & $1.735$ \\
2021 & $3.123$ & $3.063$ & $\mathllap{-}6.102$ & $0.084$ \\
2022 & $9.398$ & $9.622$ & $\mathllap{-}18.903$ & $0.117$ \\
2023 & $5.713$ & $5.254$ & $\mathllap{-}10.918$ & $0.049$ \\
\bottomrule
\end{tabular}
\endgroup

%% file: tables/regime_2017_cluster_mechanism.tex
\begingroup
\renewcommand{\arraystretch}{1.12}
\setlength{\tabcolsep}{3.8pt}
\begin{tabular}{@{}ccccccc@{}}
\toprule
Cluster & Ep. & Agent--BS $\Delta$ & $\rho(\Delta S,\Delta IV)$ & $\Delta$PnL & $\Delta$Reward & Excess Var. \\
\midrule
2017-08-03 & 30 & $\mathllap{-}0.019$ & $\mathllap{-}0.738$ & $0.117$ & $1.407$ & $35.2\%$ \\
2017-05-04 & 19 & $\mathllap{-}0.006$ & $\mathllap{-}0.705$ & $0.096$ & $1.116$ & $20.8\%$ \\
2017-03-06 & 28 & $\mathllap{-}0.028$ & $\mathllap{-}0.159$ & $0.013$ & $\mathllap{-}0.071$ & $18.2\%$ \\
2017-09-01 & 18 & $\mathllap{-}0.007$ & $0.056$ & $0.019$ & $\mathllap{-}0.053$ & $8.0\%$ \\
2017-11-02 & 18 & $\mathllap{-}0.006$ & $0.405$ & $0.060$ & $0.093$ & $7.5\%$ \\
\bottomrule
\end{tabular}
\endgroup

%% file: tables/distill_smoother_vs_agent_yearly.tex
\begingroup
\footnotesize
\renewcommand{\arraystretch}{1.16}
\setlength{\tabcolsep}{3.8pt}
\begin{tabular}{@{}cccccc@{}}
\toprule
Year & Reward & CVaR 5\% & Mean P\&L & Log Downside Variance & Log Variance \\
\midrule
2015 & $0.081$ & $\mathllap{-}0.046$ & $\mathllap{-}0.039$\sym{**} & $0.181$ & $\mathllap{-}0.046$ \\
2016 & $0.057$ & $0.014$ & $\mathllap{-}0.012$ & $0.062$ & $\mathllap{-}0.063$ \\
2017 & $0.381$\sym{***} & $0.010$ & $0.015$\sym{*} & $\mathllap{-}0.115$\sym{**} & $\mathllap{-}0.088$ \\
2018 & $0.143$\sym{*} & $0.006$ & $\mathllap{-}0.002$ & $\mathllap{-}0.016$ & $\mathllap{-}0.019$ \\
2019 & $0.328$\sym{***} & $\mathllap{-}0.003$ & $0.009$ & $\mathllap{-}0.084$ & $\mathllap{-}0.079$\sym{*} \\
2020 & $0.017$ & $\mathllap{-}0.009$ & $\mathllap{-}0.012$ & $0.023$ & $\mathllap{-}0.101$ \\
2021 & $0.119$\sym{*} & $\mathllap{-}0.013$ & $\mathllap{-}0.003$ & $0.039$\sym{*} & $0.039$ \\
2022 & $\mathllap{-}0.125$ & $\mathllap{-}0.041$\sym{***} & $\mathllap{-}0.014$\sym{*} & $0.104$\sym{***} & $0.037$ \\
2023 & $0.242$\sym{***} & $0.006$ & $\mathllap{-}0.002$ & $\mathllap{-}0.017$ & $\mathllap{-}0.069$\sym{**} \\
\bottomrule
\end{tabular}
\endgroup

%% file: tables/distill_smoother_vs_bs_summary.tex
\begingroup
\footnotesize
\renewcommand{\arraystretch}{1.16}
\setlength{\tabcolsep}{5pt}
\begin{tabular}{@{}ccccc@{}}
\toprule
Comparison & Reward & CVaR 5\% & Log Downside Variance & Log Variance \\
\midrule
Agent--BS & $0.528$ (5/1) & $0.101$ (1/0) & $\mathllap{-}0.420$ (7/0) & $0.039$ (0/1) \\
Smoothed agent--BS & $0.666$ (5/1) & $0.093$ (4/0) & $\mathllap{-}0.400$ (7/0) & $\mathllap{-}0.004$ (0/1) \\
\bottomrule
\end{tabular}
\endgroup

%% file: tables/distill_parsimonious_fit_summary.tex
\begingroup
\renewcommand{\arraystretch}{1.14}
\setlength{\tabcolsep}{4.8pt}
\begin{tabular}{@{}ccccccc@{}}
\toprule
Year & Family & Complexity & Val. MAE & Test MAE & 95\% Error & Test Corr. \\
\midrule
2015 & Smooth focus & 6 & $0.023$ & $0.023$ & $0.071$ & $0.987$ \\
2016 & Smooth uniform & 6 & $0.016$ & $0.017$ & $0.045$ & $0.993$ \\
2017 & Smooth uniform & 3 & $0.017$ & $0.036$ & $0.079$ & $0.989$ \\
2018 & Smooth uniform & 6 & $0.014$ & $0.019$ & $0.055$ & $0.996$ \\
2019 & Smooth uniform & 3 & $0.022$ & $0.023$ & $0.056$ & $0.997$ \\
2020 & Raw uniform & 10 & $0.016$ & $0.025$ & $0.074$ & $0.995$ \\
2021 & Raw uniform & 19 & $0.020$ & $0.014$ & $0.037$ & $0.999$ \\
2022 & Raw uniform & 16 & $0.010$ & $0.012$ & $0.031$ & $0.999$ \\
2023 & Smooth focus & 17 & $0.016$ & $0.020$ & $0.048$ & $0.997$ \\
\bottomrule
\end{tabular}
\endgroup

%% file: tables/distill_parsimonious_trading_summary.tex
\begingroup
\renewcommand{\arraystretch}{1.18}
\setlength{\tabcolsep}{7pt}
\begin{tabular}{@{}ccccc@{}}
\toprule
Comparison & Reward & CVaR 5\% & Log Downside Variance & Log Variance \\
\midrule
Agent--BS & $0.528$ (5/1) & $0.101$ (1/0) & $\mathllap{-}0.420$ (7/0) & $0.039$ (0/1) \\
Formula--BS & $1.131$ (6/1) & $0.123$ (6/0) & $\mathllap{-}0.518$ (8/0) & $0.013$ (0/1) \\
Formula--Agent & $0.604$ (4/1) & $0.022$ (1/0) & $\mathllap{-}0.098$ (1/1) & $\mathllap{-}0.026$ (1/1) \\
\bottomrule
\end{tabular}
\endgroup

%% file: tables/robust_selection_rule_summary.tex
\begingroup
\renewcommand{\arraystretch}{1.16}
\setlength{\tabcolsep}{3.8pt}
\begin{tabular}{@{}cccc@{\hspace{8pt}}ccc@{}}
\toprule
Rule & Comp. & MAE & Rew--BS & Down--BS & Rew--Agent & Down--Agent \\
\midrule
Pars. 10\% & $9.6$ & $0.021$ & $1.131$ (6/1) & $-0.518$ (8/0) & $0.604$ (4/1) & $-0.098$ (1/1) \\
Val. fit & $16.6$ & $0.020$ & $1.131$ (6/1) & $-0.493$ (8/0) & $0.603$ (4/0) & $-0.074$ (1/1) \\
Val. reward & $7.2$ & $0.027$ & $1.133$ (6/1) & $-0.526$ (7/0) & $0.605$ (3/1) & $-0.106$ (2/1) \\
Val. CVaR & $11.1$ & $0.023$ & $0.954$ (6/1) & $-0.502$ (8/0) & $0.426$ (4/1) & $-0.083$ (1/1) \\
Val. downside & $5.9$ & $0.028$ & $1.022$ (5/1) & $-0.544$ (7/0) & $0.494$ (4/1) & $-0.125$ (3/1) \\
\bottomrule
\end{tabular}
\endgroup

%% file: tables/robust_hull_white_summary.tex
\begingroup
\renewcommand{\arraystretch}{1.18}
\setlength{\tabcolsep}{7pt}
\begin{tabular}{@{}ccccc@{}}
\toprule
Comparison & Reward & CVaR 5\% & Log Downside Variance & Log Variance \\
\midrule
Hull-White vs. Agent & $\mathllap{-}0.144$ (2/2) & $\mathllap{-}0.130$ (0/3) & $0.246$ (0/3) & $\mathllap{-}0.044$ (1/1) \\
Hull-White vs. Formula & $\mathllap{-}0.747$ (0/2) & $\mathllap{-}0.152$ (0/2) & $0.344$ (0/4) & $\mathllap{-}0.018$ (2/2) \\
\bottomrule
\end{tabular}
\endgroup

%% file: tables/robust_seed_metrics_new_seed_final_WF_exp1_k1_test.tex
\begingroup
\renewcommand{\arraystretch}{1.15}
\setlength{\tabcolsep}{5pt}
\begin{tabular}{@{}cccccc@{}}
\toprule
Year & Reward & CVaR 5\% & Mean P\&L & Log Downside Variance & Log Variance \\
\midrule
2015 & $1.875$\sym{***} & $0.252$\sym{***} & $0.129$\sym{**} & $\mathllap{-}0.730$\sym{***} & $0.050$ \\
2016 & $1.149$\sym{***} & $0.165$\sym{*} & $0.067$\sym{*} & $\mathllap{-}0.459$\sym{***} & $0.002$ \\
2017 & $0.322$\sym{**} & $0.007$ & $0.030$\sym{***} & $\mathllap{-}0.089$\sym{**} & $0.152$\sym{**} \\
2018 & $0.595$ & $0.147$ & $0.079$\sym{**} & $\mathllap{-}0.409$\sym{*} & $0.112$ \\
2019 & $0.989$ & $0.017$ & $0.124$\sym{**} & $\mathllap{-}0.473$\sym{**} & $\mathllap{-}0.121$ \\
2020 & $\mathllap{-}0.031$ & $0.094$ & $0.281$\sym{**} & $\mathllap{-}0.670$\sym{***} & $0.185$ \\
2021 & $1.406$\sym{***} & $0.063$\sym{***} & $0.107$\sym{***} & $\mathllap{-}0.404$\sym{***} & $0.126$ \\
2022 & $\mathllap{-}3.199$\sym{***} & $\mathllap{-}0.034$ & $\mathllap{-}0.045$ & $0.162$ & $\mathllap{-}0.017$ \\
2023 & $0.275$ & $0.048$ & $0.089$\sym{***} & $\mathllap{-}0.173$\sym{*} & $0.523$\sym{***} \\
\bottomrule
\end{tabular}
\endgroup

%% file: tables/robust_seed_metrics_1_seed_final_WF_exp1_k1_test.tex
\begingroup
\renewcommand{\arraystretch}{1.15}
\setlength{\tabcolsep}{5pt}
\begin{tabular}{@{}cccccc@{}}
\toprule
Year & Reward & CVaR 5\% & Mean P\&L & Log Downside Variance & Log Variance \\
\midrule
2015 & $1.779$\sym{***} & $0.253$ & $0.135$\sym{*} & $\mathllap{-}0.835$\sym{***} & $\mathllap{-}0.199$ \\
2016 & $1.352$\sym{***} & $0.125$ & $0.071$\sym{*} & $\mathllap{-}0.453$\sym{***} & $0.030$ \\
2017 & $0.503$\sym{***} & $0.018$ & $0.033$\sym{***} & $\mathllap{-}0.124$\sym{**} & $0.112$ \\
2018 & $0.373$ & $0.167$ & $0.049$ & $\mathllap{-}0.418$\sym{*} & $0.043$ \\
2019 & $1.625$\sym{***} & $0.110$\sym{*} & $0.139$\sym{**} & $\mathllap{-}0.668$\sym{***} & $\mathllap{-}0.215$ \\
2020 & $\mathllap{-}0.876$ & $\mathllap{-}0.120$ & $0.251$ & $\mathllap{-}0.331$ & $0.096$ \\
2021 & $1.682$\sym{***} & $0.088$\sym{***} & $0.117$\sym{***} & $\mathllap{-}0.492$\sym{***} & $0.084$ \\
2022 & $\mathllap{-}1.969$\sym{***} & $\mathllap{-}0.019$ & $\mathllap{-}0.018$ & $\mathllap{-}0.005$ & $\mathllap{-}0.067$ \\
2023 & $\mathllap{-}0.257$ & $0.041$ & $0.106$\sym{***} & $\mathllap{-}0.164$ & $0.682$\sym{***} \\
\bottomrule
\end{tabular}
\endgroup

%% file: tables/robust_hull_white_vs_bs.tex
\begingroup
\renewcommand{\arraystretch}{0.95}
\setlength{\tabcolsep}{4.5pt}
\begin{tabular}{@{}cccccc@{}}
\toprule
Year & Reward & CVaR 5\% & Mean P\&L & Log Downside Variance & Log Variance \\
\midrule
2015 & $\mathllap{-}4.236$\sym{***} & $\mathllap{-}0.264$\sym{***} & $\mathllap{-}0.237$\sym{***} & $0.612$\sym{***} & $\mathllap{-}0.186$ \\
2016 & $0.121$ & $\mathllap{-}0.119$ & $\mathllap{-}0.040$ & $0.169$ & $0.026$ \\
2017 & $1.015$\sym{**} & $\mathllap{-}0.030$ & $0.077$\sym{***} & $\mathllap{-}0.178$ & $0.452$\sym{*} \\
2018 & $1.202$\sym{***} & $0.115$ & $\mathllap{-}0.030$ & $\mathllap{-}0.386$\sym{*} & $\mathllap{-}0.241$\sym{**} \\
2019 & $2.441$\sym{***} & $0.186$\sym{**} & $0.232$\sym{**} & $\mathllap{-}1.186$\sym{***} & $\mathllap{-}0.356$ \\
2020 & $2.860$\sym{**} & $\mathllap{-}0.087$ & $0.112$ & $\mathllap{-}0.482$ & $\mathllap{-}0.494$ \\
2021 & $2.188$\sym{***} & $0.058$ & $0.174$\sym{***} & $\mathllap{-}0.477$\sym{***} & $0.460$\sym{**} \\
2022 & $\mathllap{-}2.303$\sym{***} & $\mathllap{-}0.149$ & $\mathllap{-}0.105$ & $0.446$ & $0.027$ \\
2023 & $0.171$ & $0.031$ & $0.039$\sym{*} & $\mathllap{-}0.076$ & $0.271$\sym{***} \\
\bottomrule
\end{tabular}
\endgroup

%% file: tables/robust_hull_white_vs_agent.tex
\begingroup
\renewcommand{\arraystretch}{0.95}
\setlength{\tabcolsep}{4.5pt}
\begin{tabular}{@{}cccccc@{}}
\toprule
Year & Reward & CVaR 5\% & Mean P\&L & Log Downside Variance & Log Variance \\
\midrule
2015 & $\mathllap{-}6.182$\sym{***} & $\mathllap{-}0.501$\sym{***} & $\mathllap{-}0.386$\sym{***} & $1.414$\sym{***} & $\mathllap{-}0.009$ \\
2016 & $\mathllap{-}1.181$\sym{***} & $\mathllap{-}0.255$\sym{***} & $\mathllap{-}0.101$\sym{*} & $0.598$\sym{**} & $0.057$ \\
2017 & $0.606$ & $\mathllap{-}0.067$ & $0.032$ & $\mathllap{-}0.017$ & $0.266$ \\
2018 & $0.679$ & $\mathllap{-}0.052$ & $\mathllap{-}0.084$\sym{*} & $0.081$ & $\mathllap{-}0.255$\sym{**} \\
2019 & $1.236$\sym{**} & $0.084$ & $0.114$\sym{*} & $\mathllap{-}0.630$\sym{*} & $\mathllap{-}0.170$ \\
2020 & $3.358$\sym{***} & $\mathllap{-}0.133$ & $\mathllap{-}0.196$\sym{***} & $0.195$ & $\mathllap{-}0.485$\sym{*} \\
2021 & $0.514$ & $\mathllap{-}0.024$ & $0.054$\sym{***} & $\mathllap{-}0.026$ & $0.317$\sym{**} \\
2022 & $\mathllap{-}0.398$ & $\mathllap{-}0.196$\sym{***} & $\mathllap{-}0.089$\sym{***} & $0.508$\sym{***} & $0.144$ \\
2023 & $0.076$ & $\mathllap{-}0.026$ & $\mathllap{-}0.048$\sym{***} & $0.092$ & $\mathllap{-}0.256$\sym{*} \\
\bottomrule
\end{tabular}
\endgroup

%% file: tables/robust_hull_white_vs_formula.tex
\begingroup
\renewcommand{\arraystretch}{0.95}
\setlength{\tabcolsep}{4.5pt}
\begin{tabular}{@{}cccccc@{}}
\toprule
Year & Reward & CVaR 5\% & Mean P\&L & Log Downside Variance & Log Variance \\
\midrule
2015 & $\mathllap{-}6.707$\sym{***} & $\mathllap{-}0.586$\sym{***} & $\mathllap{-}0.434$\sym{***} & $1.714$\sym{***} & $\mathllap{-}0.120$ \\
2016 & $\mathllap{-}1.285$\sym{***} & $\mathllap{-}0.241$\sym{*} & $\mathllap{-}0.087$ & $0.496$\sym{**} & $0.109$ \\
2017 & $\mathllap{-}0.166$ & $\mathllap{-}0.058$ & $\mathllap{-}0.002$ & $0.159$ & $0.289$\sym{**} \\
2018 & $\mathllap{-}0.013$ & $\mathllap{-}0.038$ & $\mathllap{-}0.129$\sym{**} & $0.103$ & $\mathllap{-}0.353$\sym{***} \\
2019 & $0.220$ & $\mathllap{-}0.071$ & $0.027$ & $\mathllap{-}0.051$ & $\mathllap{-}0.133$ \\
2020 & $0.911$ & $\mathllap{-}0.157$ & $\mathllap{-}0.159$\sym{***} & $0.285$\sym{*} & $\mathllap{-}0.217$ \\
2021 & $0.478$ & $\mathllap{-}0.036$ & $0.061$\sym{***} & $\mathllap{-}0.006$ & $0.355$\sym{***} \\
2022 & $0.081$ & $\mathllap{-}0.138$\sym{***} & $\mathllap{-}0.049$\sym{***} & $0.278$\sym{***} & $0.095$\sym{*} \\
2023 & $\mathllap{-}0.244$ & $\mathllap{-}0.040$ & $\mathllap{-}0.051$\sym{***} & $0.121$\sym{***} & $\mathllap{-}0.183$\sym{**} \\
\bottomrule
\end{tabular}
\endgroup

%% file: tables/robust_haircut_summary.tex
\begingroup
\renewcommand{\arraystretch}{1.05}
\setlength{\tabcolsep}{7pt}
\begin{tabular}{@{}ccccc@{}}
\toprule
Comparison & Reward & CVaR 5\% & Log Downside Variance & Log Variance \\
\midrule
Haircut--BS & $0.186$ (5/2) & $-0.074$ (1/1) & $0.030$ (2/1) & $-0.080$ (1/1) \\
Haircut--Agent & $-0.341$ (0/5) & $-0.175$ (0/3) & $0.449$ (0/6) & $-0.119$ (1/1) \\
Haircut--Formula & $-0.945$ (0/5) & $-0.196$ (0/7) & $0.547$ (0/9) & $-0.093$ (2/1) \\
\bottomrule
\end{tabular}
\endgroup

%% file: tables/robust_switching_negative_result.tex
\begingroup
\renewcommand{\arraystretch}{0.95}
\setlength{\tabcolsep}{3.8pt}
\begin{tabular}{@{}ccccccc@{}}
\toprule
Policy & MAE & 95\% err. & Rew--Agent & Down--Agent & Rew--BS & Down--BS \\
\midrule
Global & $0.012$ & $0.031$ & $\mathllap{-}0.479$\sym{**} & $0.230$\sym{**} & $\mathllap{-}2.384$\sym{**} & $0.167$ \\
2-band & $0.014$ & $0.037$ & $\mathllap{-}0.402$\sym{**} & $0.287$\sym{**} & $\mathllap{-}2.307$\sym{**} & $0.224$ \\
3-band & $0.011$ & $0.031$ & $\mathllap{-}0.251$ & $0.180$\sym{**} & $\mathllap{-}2.156$\sym{**} & $0.117$ \\
\bottomrule
\end{tabular}
\endgroup